%% file: main.tex
\documentclass[10pt]{article}
\usepackage{geometry}                
\usepackage[T1]{fontenc}
\usepackage{graphicx}
\usepackage{amssymb}
\usepackage{amsmath}
\usepackage{mathtools}
\usepackage{bm}
\usepackage{siunitx}
\usepackage{hyperref}
\hypersetup{
    hidelinks,
    colorlinks=true,
    }
\usepackage{xcolor}
\usepackage{authblk}
\usepackage[numbers]{natbib}
\usepackage{siunitx}
\usepackage[normalem]{ulem}

\usepackage{tikz}
\usepackage{pgfplots}
\usepackage{subcaption}
\usepackage{caption}
\usepackage{algorithm}
\usepackage{algpseudocode}
\usepackage{multirow}
\usepackage{makecell}
\usepackage{rotating}
\usepackage{bigstrut}

\usepackage{array}

\pgfplotsset{width=10cm,compat=1.9}

\input{aeromacros.tex}
\input{mycolors.tex}

\DeclareMathAlphabet{\mathsfi}{\encodingdefault}{\sfdefault}{m}{sl}
\SetMathAlphabet{\mathsfi}{bold}{\encodingdefault}{\sfdefault}{bx}{sl}

\begin{document}

\title{Deep reinforcement learning of airfoil pitch control in a highly disturbed environment using partial observations}
\author[1]{Diederik Beckers\thanks{Email address for correspondence: beckers@caltech.edu}}
\author[2]{Jeff D. Eldredge}
\affil[1]{Graduate Aerospace Laboratories, California Institute of Technology, Pasadena, CA 91125 USA}
\affil[2]{Mechanical and Aerospace Engineering, University of California, Los Angeles, CA 90095-1597 USA}
\date{}
\maketitle

\begin{abstract}
    This study explores the application of deep reinforcement learning (RL) to design an airfoil pitch controller capable of minimizing lift variations in randomly disturbed flows. The controller, treated as an agent in a partially observable Markov decision process, receives non-Markovian observations from the environment, simulating practical constraints where flow information is limited to force and pressure sensors. Deep RL, particularly the TD3 algorithm, is used to approximate an optimal control policy under such conditions. Testing is conducted for a flat plate airfoil in two environments: a classical unsteady environment with vertical acceleration disturbances (i.e., a Wagner setup) and a viscous flow model with pulsed point force disturbances. In both cases, augmenting observations of the lift, pitch angle, and angular velocity with extra wake information (e.g., from pressure sensors) and retaining memory of past observations enhances RL control performance. Results demonstrate the capability of RL control to match or exceed standard linear controllers in minimizing lift variations. Special attention is given to the choice of training data and the generalization to unseen disturbances.
\end{abstract}

\section{Introduction}

Large flow disturbances can drastically alter the aerodynamic forces on an airfoil and even induce flow separation, possibly compromising the safety, performance, and lifetime of aerodynamic vehicles and energy-conversion systems. Flow disturbances can appear as localized flow features, or gusts, that are incident on the airfoil, either as discrete instances or as continuous variations, or they can manifest as unsteady ambient flow, for instance, a varying freestream flow. The interaction of large-amplitude flow disturbances with an airfoil and the flow response to actuation is generally characterized by the nonlinear formation and shedding of leading- and trailing-edge vortices and is not always fully understood for every type of actuation or disturbance. Furthermore, the flow response during a rapid succession of such actuations and disturbances cannot be simply superposed~\citep{an2016}, and their large parameter spaces~\citep{jones2020} complicate the design of a comprehensive control strategy. This is further complicated by the limited availability of sensor information in practical applications.

Given the control objectives (e.g., lift enhancement, drag reduction, stall delay, etc.), the actuation method (e.g., pitching, control surface actuation, synthetic jets, plasma actuators, etc.), and the available measurements (e.g., pressure, velocity, force, torque, etc.), the control of unsteady flows over airfoils requires the selection of a specific control strategy and design approach. Active control strategies can be predetermined, feedforward, or feedback control, and the design approach can be signal-based, frequency-based, or based on numerical optimization~\citep{skogestad2005}. In the domain of unsteady aerodynamics, common control systems target the reduction of lift variations due to unstable states or disturbances using feedback control. Signal-based approaches to such control systems assume that the unstable behavior or the shape or time profile of the disturbance is known, and examples include the use of a linear quadratic regulator to suppress vortex shedding using point forcing near the trailing edge~\citep{ahuja2010} and the lift regularization for a flat plate encountering a transverse gust through pitching based on heuristics~\citep{sedky2022}. Frequency-based control aims first to shape the open-loop or closed-loop transfer functions to achieve specific robustness or performance objectives and then design a controller whose transfer functions approximate those shapes. Examples of such control for lift mitigation in unsteady aerodynamics include controller designs based on $\mathcal{H}_{\infty}$-synthesis to reject vertical acceleration disturbances for a flat plate through pitching~\citep{brunton2013,brunton2014,sedky2020} and to reject longitudinal gusts for semi-circular wings using pulsed blowing~\citep{kerstens2011}.

The previous examples employed linear controllers that are designed using a linearized model of the flow. However, for large flow disturbances or a rapid succession of disturbances and actuation, modeling and controlling nonlinear effects become increasingly important. Simultaneously, the extensive parameter spaces associated with disturbances and actuation impede the practical application of linearized models. Nonlinear controllers can be designed using the numerical optimization approach, which is witnessing rapid advancement due to contributions emerging from the machine learning and data science domains. Notable flow control examples include model predictive control of cylinder rotation to control the wake~\citep{bieker2020}, iterative maneuver optimization to rapidly design predetermined nonlinear pitch control for lift regulation during gust encounters~\citep{xu2023}, and reinforcement learning, which we discuss next.

Reinforcement learning solves control problems formulated as Markov decision processes by iteratively learning optimal control through trial-and-error interactions with an environment. In its early applications, reinforcement learning (RL) relied on tabular representations for the value of every possible state in a Markov decision process to optimize a control law, or policy, which is only tractable for modest-dimensional, discrete problems. In the past decade, however, the successful integration of deep learning into RL (forming deep RL) enabled the approximation of policies and value functions over high-dimensional continuous spaces using multi-layered artificial neural networks, leading to many breakthroughs in artificial intelligence, especially in the fields of autonomous systems and controls~\citep{arulkumaran2017}. Furthermore, the use of policy and value function approximations extends RL to solving partially observable Markov decision processes (POMDP)~\citep{sutton2018,kurniawati2022}, i.e., Markov decision processes where the full state of the system cannot be directly measured and whose exact solutions would be intractable except for the smallest problems. Thanks to these features, deep RL has been successfully applied to the control or modeling of high-dimensional fluid flows with uncertainty about the underlying state, e.g. navigating through wakes~\citep{verma2018,novati2019,gunnarson2021} or vortex shedding suppression~\citep{fan2020,nair2023,xia2024} with limited sensor information, or uncertainty about model parameters, e.g. from turbulence models~\citep{novati2021,bae2022}. Similar to the objective of the previously introduced signal- and frequency-based control problems, \citet{renn2022} apply deep RL to learn airfoil flap control to minimize lift variations on an airfoil positioned downstream of a vortex-shedding cylinder using only lift measurements.

\begin{figure}
    \centering
    \includegraphics[width=0.75\linewidth]{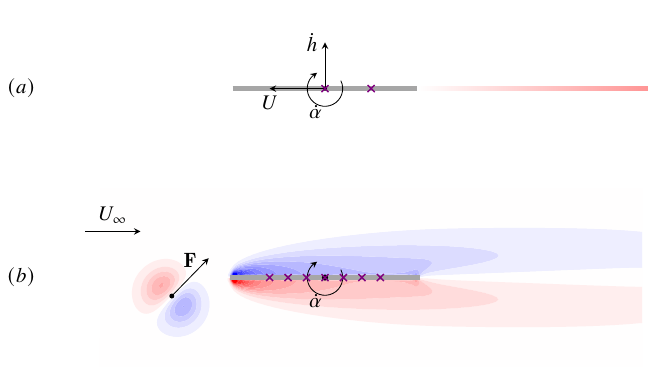} 
    \caption{$(a)$ Diagram of the airfoil's motion and wake for the classical unsteady aerodynamics environment. $(b)$ Diagram of the airfoil, wake, freestream, and forcing for the viscous flow environment. The crosses indicate the positions of the pressure sensors, when used.}
    \label{fig:combined_diagram}
\end{figure}

In this work, we apply deep RL for learning airfoil pitch control to minimize lift variations due to flow disturbances in two types of flow environments that model the flow around the airfoil, demonstrated in Figure~\ref{fig:combined_diagram}: one type that models the flow purely as a potential flow with small disturbances, thus allowing us to simplify the system to a classical aerodynamics setup with relatively few states, and another type that simulates the viscous flow at a low Reynolds number with large disturbances, necessitating the use of a high-dimensional model of this nonlinear system. We use these environments to focus on two aspects of this reinforcement learning problem: the partial observability, or how the performance of the control depends on the type and availability of measurements of the state of the system; and the generalizability of the trained controller, or how a policy trained on one type of disturbance performs when faced with different types of disturbances. To apply reinforcement learning to the control problem, we first formulate it as a (partially observable) Markov decision process and select an appropriate learning algorithm, described in Section~\ref{sec:control_framework}. We then analyze the performance of the RL pitch control using different training setups in an environment with the classical unsteady aerodynamics model in Section~\ref{sec:classical_environment} and compare it with a linear-quadratic-Gaussian controller for smooth and impulsive vertical acceleration disturbances. In Section~\ref{sec:viscous_environment}, we perform a similar analysis of the RL performance in an environment with the viscous flow model for flow disturbances introduced upstream of the airfoil with a pulsed point force.

\section{The pitch control learning framework} \label{sec:control_framework}

Consider a flat plate airfoil oriented at an angle $\AOA$ with the negative $\coordinertial{x}$-axis of an inertial reference frame (positive in the clockwise direction). We describe the kinematics of the airfoil in terms of the velocity of a pivot point on the chord and the angular velocity $\AOAdot$ about this pivot point. We restrict the horizontal motion of the pivot point to a constant velocity with magnitude $\pivotxvel$ in the negative $\coordinertial{x}$ direction. Equivalently for a flat plate, we can also analyze the airfoil from the perspective of a reference frame with the same orientation as the inertial reference frame, centered at and moving with the pivot point. In this specific moving reference frame, which we will refer to as the windtunnel reference frame, the flow around the airfoil from its horizontal motion is modeled with a constant horizontal freestream velocity $\freestreamxvel=\pivotxvel$. In this work, we focus on the control of the airfoil's pitch by varying the angular acceleration $\AOAddot$ about the pivot point while the airfoil encounters disturbances that affect the aerodynamic forces that act on it. The specific disturbances we will study consist of randomized uniform vertical flow accelerations $\pivotyaccel$ or vortex structures generated by a randomized forcing $\forcefield$ in the flow upstream of the airfoil. A set of measurements is fed back to the controller. These can consist of the angle $\AOA$, angular velocity $\AOAdot$, lift $\lift$, pressure $\pressure$, and the states encoding the wake evolution in the Wagner model, with the exact combination of measurements depending on the case.

\begin{figure}
    \centering
    \includegraphics[width=0.6\linewidth]{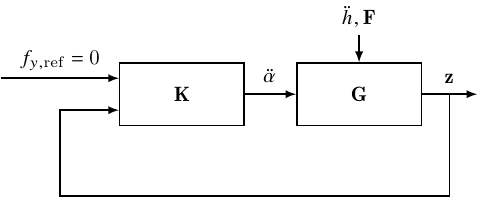} 
    \caption{Control diagram for the pitch control problem.}
    \label{fig:control_problem}
\end{figure}

The control problem is to design the discrete-time feedback controller $\TFgeneralcontroller$ in Figure~\ref{fig:control_problem} that pitches the airfoil such that lift variations about a prescribed reference are minimized and $\AOA$, $\AOAdot$, and $\AOAddot$ are limited to prescribed ranges. In this work, we set the reference lift to zero, but the so-called servo problem of tracking other reference lifts can be explored by straightforward extension.

\subsection{The pitch control problem formulated as a POMDP}

\begin{figure}
    \centering
    \includegraphics{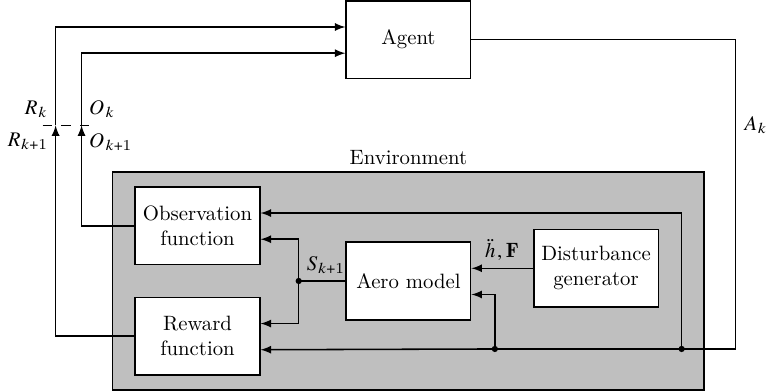} 
    \caption{Diagram of the POMDP learning framework for the pitch control problem.}
    \label{fig:diagram_RL}
\end{figure}

We can formulate the control problem as a POMDP framework~\citep{kaelbling1998,sutton2018} following the agent-environment structure depicted in Figure~\ref{fig:diagram_RL}. Here, the agent assumes the role of the controller, and the environment contains all the processes that the agent cannot control arbitrarily, i.e., the aerodynamics model, the disturbance generator, and the reward and observation functions. If we were to consider a non-zero reference lift, the process that generates this signal would also have to be included in the environment. The aerodynamics model and the disturbance and reference lift generators together define the probabilistic state transition function that, during each timestep $\currenttimestep$, advances the environment state $\MDPstatevar[\currenttimestep]$ to $\MDPstatevar[\currenttimestep+1]$ as a result of the latest action $\MDPactionvar[\currenttimestep] \coloneqq \AOAddot[\currenttimestep]$ of the agent. This action remains piecewise constant over the time between two agent-environment interactions. The agent then receives a partial observation $\POMDPobservationvar[\currenttimestep+1]$ of the environment and a reward $\MDPrewardvar[\currenttimestep+1]$ resulting from that action, which it maps to its next action $\MDPactionvar[\currenttimestep+1]$. The POMDP formalizes the sequential decision making process consisting of the sequence $\POMDPobservationvar[1],\MDPactionvar[1],\ldots,\POMDPobservationvar[\currenttimestep], \MDPactionvar[\currenttimestep], \MDPrewardvar[\currenttimestep+1], \POMDPobservationvar[\currenttimestep+1],\ldots$. For simplicity, we only study sequences of finite length, with each such sequence called an \emph{episode}. Solving the POMDP amounts to finding the optimal policy that maximizes the expected sum of rewards in one episode, also called the episode \emph{return}, thus finding a decision-making strategy that selects actions accounting for potential long-term rewards but also uncertainty about the environment.

\begin{figure}
    \centering
    \includegraphics{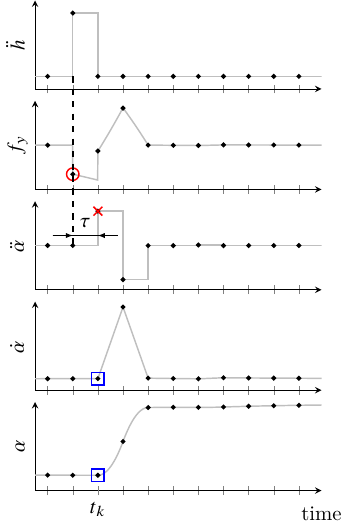}
    \caption{Behavior of the lift $\lift$, control input $\AOAddot$, and angular states $\AOAdot$ and $\AOA$ during the flow response to an impulsive $\pivotyaccel$ disturbance with output feedback control. The control input is piecewise constant between timesteps and the pitching is applied at the midchord point. The lift output observed by the controller is delayed, in this case by the same time as the action timestep, and the feedback controller's action at timestep $\currenttimestep$ (\ref{ls:control_decision}) is therefore based on the the lift from the previous timestep (\ref{ls:lift_measurement}). In contrast, for the control decision at that same timestep, all the RL agents in this work additionally observe the angular states at $\currenttimestep$ (\ref{ls:angular_measurement}) and potentially the pressure or other wake information (not shown).}
    \label{fig:impulse_response}
\end{figure}

The state $\MDPstatevar[\currenttimestep]$ of the environment is composed of the state of the aerodynamics model at timestep $\currenttimestep$, which in our work comprises the kinematic state of the airfoil; the state of the flow around it, which can be stochastically altered by the disturbance generator; the previous action $\MDPactionvar[\currenttimestep-1]$ for use in the reward function; and any variables needed to construct the elements of the measured variables in $\measurement$. The initial state of the aerodynamic model at the beginning of every episode is the steady-state motion of the airfoil at zero angle of attack and the corresponding steady-state flow solution. We will impose bounds on the environment's state for training and stability purposes. If these bounds are violated at any point during the episode, the episode is terminated early and reset to the initial conditions. 

The observation function maps the state and action to an observation for the agent. In our model, we assume that this is done without additional noise. For the cases in this work, $\POMDPobservationvar[\currenttimestep]$ always contains $\AOA[\currenttimestep]$, $\AOAdot[\currenttimestep]$, and the latest available lift $\lift$. Since the lift generally depends on the action at the same instant (through the inertial reaction of the fluid), we assume that the agent only has knowledge of the lift from an earlier time instant, creating a delay in the response to a disturbance, as illustrated in Figure~\ref{fig:impulse_response}. We will assume that this output delay is also present in other control strategies that we use for comparison in this paper. Though this delay limits the controller's performance, it is likely to arise in practice. Depending on the studied case, the observation vector may also include the latest available pressure values at the airfoil or, in one case, the full state of the flow, all from the same time instant as the lift. 

The final element of the environment is the reward function, which maps the new state $\MDPstatevar[\currenttimestep+1]$ and the last action $\MDPactionvar[\currenttimestep]$ to a reward signal $\MDPrewardvar[\currenttimestep+1]$, used in the objective function to optimize the policy. For this reason, the reward function should introduce the control objective into the POMDP framework. 

\subsection{The agent and the learning algorithm}

The agent's task is to find the optimal policy for its actions, leveraging the rewards and observations it receives following each action for learning this policy. In the online reinforcement learning approach, it considers the environment as a black box and learns about its behavior while simultaneously interacting with it. To balance between exploiting the learned policy to maximize rewards and exploring new actions to discover potentially better strategies, the agent often incorporates stochastic behavior during its interactions with the environment. 

In a realistic setting, the state of the environment is continuous-valued and usually highly dimensional due to the flow dynamics. The observations rarely encompass the entire flow field, rendering the environment's state only partially observable and resulting in the agent receiving non-Markovian signals. In the POMDP formalism, the agent addresses this uncertainty by maintaining a belief state -- a probability distribution reflecting the true state of the environment~\citep{kaelbling1998}. This belief state informs optimal decision-making, with the agent employing a policy that maps these belief states to actions. Alternatively, effective strategies can take a more straightforward approach, partially restoring the Markov property by utilizing a fixed-width finite history window of recent observations fed to a neural network or a recurrent neural network~\citep{kaelbling1996,lin1992}. These memory-based approaches to POMDPs have been applied in recent years with great success using model-free deep reinforcement learning techniques~\citep{mnih2013,hausknecht2015,meng2021}, and we adopt here the deep reinforcement learning approach of using a fixed-width finite history window of recent observations.

Model-free approaches require sampling many interactions with the environment, which can be computationally expensive for high-fidelity flow simulators. This warrants the use of an off-policy learning algorithm, which makes more efficient use of each sampled interaction than on-policy algorithms. Because we also desire a control law for a continuous-valued action, the training uses the twin delayed deep deterministic policy gradient (TD3) algorithm~\citep{fujimoto2018}, which we found in previous work~\citep{phdthesis} to perform better in this problem than alternative state-of-the-art, off-policy, continuous-action algorithms, such as soft actor-critic~\citep{haarnoja2018}. TD3 employs a deterministic policy and stochastically explores actions and states during training by adding normally-distributed \emph{action noise} to the actions chosen by the latest policy. Details of the algorithm's implementation in this work can be found in Appendix~\ref{app:rl_details}.

\section{Learning in a classical unsteady aerodynamics environment} \label{sec:classical_environment}

In this first example of the control analysis, we consider an inviscid, two-dimensional incompressible flow and disturbances that are either random vertical accelerations $\pivotyaccel$ of the plate's pivot point or, equivalently for a flat plate, random accelerations of a vertical freestream flow. If we assume the resulting velocities from the disturbance and pitch control are small compared to the forward velocity, i.e., $\pivotyvel \ll \pivotxvel$ and $\AOAdot \chord \ll \pivotxvel$, then we can assume that the wake's vorticity is concentrated in an infinitesimally-thin vortex sheet that convects away from the plate's trailing edge with the relative velocity $\pivotxvel$ and remains parallel to the plate, illustrated in the top panel of Figure~\ref{fig:combined_diagram}. Under these assumptions, the theory of classical unsteady aerodynamics as analyzed by \citet{wagner1925} and \citet{theodorsen1935} is valid, and the following linear dependence of the lift coefficient $\liftcoef = 2 \lift / \left( \rho \pivotxvel^2 \chord \right)$ on the disturbance $\pivotyaccel$ and control input $\AOAddot$ can be derived (see Appendix~\ref{app:classical_aero}):

\begin{gather}
    \label{eq:text_jones_ss_state}
    \begin{bmatrix}
        \AOAeffdot \\
        \AOAddot \\
        \theoboldstatedot
    \end{bmatrix}
    =
    \begin{bmatrix}
        0 & 1 & \vect{0} \\
        0 & 0 & \vect{0} \\
        2 \pi \sstheoB & \vect{0} & \sstheoA
    \end{bmatrix}
    \begin{bmatrix}
        \AOAeff \\
        \AOAdot \\
        \theoboldstate
    \end{bmatrix}
    +
    \begin{bmatrix}
        -\frac{1}{\pivotxvel} & \frac{1}{\pivotxvel} \left( \frac{\chord}{4} - \pivotdist \right) \\
        0 & 1 \\
        \vect{0} & \vect{0}
    \end{bmatrix}
    \begin{bmatrix}
        \pivotyaccel \\
        \AOAddot 
    \end{bmatrix}
    \\
    \label{eq:text_jones_ss_output}
    \liftcoef
    =
    \begin{bmatrix}
        2 \pi \sstheoD & \frac{\pi \chord}{2 \pivotxvel} & \sstheoC
    \end{bmatrix}
    \vphantom{\begin{bmatrix}
        \AOAeff \\
        \AOAdot \\
        \theoboldstate
    \end{bmatrix}
    }
    \begin{bmatrix}
        \AOAeff \\
        \AOAdot \\
        \theoboldstate
    \end{bmatrix}
    +
    \begin{bmatrix}
        -\frac{\pi \chord}{2 \pivotxvel^2} & -\frac{\pi \chord \pivotdist}{2 \pivotxvel^2}
    \end{bmatrix}
    \vphantom{\begin{bmatrix}
        \AOAeff \\
        \AOAdot \\
        \theoboldstate
    \end{bmatrix}
    }
    \begin{bmatrix}
        \pivotyaccel \\
        \AOAddot 
    \end{bmatrix} \quad.
\end{gather}
Here, $\AOAeff = -\pivotyvel/\pivotxvel + \AOA + \left( \chord/4 - \pivotdist \right) \AOAdot/\pivotxvel$ is the effective angle of attack, and ($\sstheoA$, $\sstheoB$, $\sstheoC$, $\sstheoD$) represents the state-space system that models the wake dynamics using the 2-element state vector $\theoboldstate$. We will only analyze cases where the pivot point is the midchord point, so $\pivotdist=0$. We note that, in this configuration, the angular acceleration has no direct contribution to the lift and the lift value could in principle inform the action value at the same timestep. However, for generalizability purposes and to create a more challenging and realistic control problem, we let the agent observe the lift with a time delay $\controldelay$ equal to the agent-interaction timestep.

\subsection{Setup}
From now on, we non-dimensionalize variables using $\pivotxvel$, $\chord$, and $\rho$, and the reader should assume that any variable is dimensionless. The timestep $\tau$ between every agent-environment interaction corresponds to 0.1 convective times and we simulate episodes of at most 200 timesteps with the discrete-time form of the state-space system~\eqref{eq:text_jones_ss_state}-\eqref{eq:text_jones_ss_output}. For later purposes, we define a scale $\scalepivotyaccel=0.01$ for the disturbance and  $\scaleliftcoef=0.02$ for the lift coefficient. We use a value $\maxAOAddot=0.1$ to limit the control input to the bounds $[-\maxAOAddot, \maxAOAddot]$, which is large enough to avoid saturation while achieving the control objective, except for during the early stages of the training. However, if the action space were smaller, the nonlinear nature of deep reinforcement learning is expected to deal adequately with the nonlinear saturation, which might not always be possible with linear control theory. 

We study the response to two types of $\pivotyaccel$ variations, which we will label respectively as impulsive and smooth and which do not occur simultaneously in the same episode. The value of $\pivotyaccel$ over the period between two action timesteps is always piecewise constant. The smooth disturbances are constructed from a Fourier sine series containing modes 5 through 50, with 2.5 to 25 cycles in one episode, respectively. The $n$th mode has an amplitude $D/n$ with $D$ sampled from a standard normal distribution for each mode. This choice ensures that the variations in $\pivotyaccel$ are neither too slow, which would lead to high $\pivotyvel$, nor too fast for the controller to react. The signal is then scaled such that the maximum amplitude is equal to $\scalepivotyaccel$. The impulsive disturbances consist of a random number, uniformly sampled from $[0,20]$, of events, occurring at unique random discrete times between timesteps 0 and 199. An event has equal likelihood of being either a step change or a discrete (finite-amplitude) impulse. The amplitude of each event is randomly chosen from a uniform distribution with a zero lower bound and an upper bound that is limited such that $|\pivotyvel| < 0.01$ and $|\pivotyaccel| < 0.7 \scalepivotyaccel$, chosen to approximately match the uncontrolled $\liftcoef$ magnitudes from the smooth disturbances.

During training, we terminate the episode early if $|\liftcoef| > \scaleliftcoef$ to expedite the learning process; this reduces the computational time spent on simulating high-lift states that are reached after poor pitch control. We do not restrict $\AOAdot$ or $\AOA$, although this could also be used to optimize the training, as we will do in the viscous problem in the next section. To train the agent to minimize the lift variations, the reward after every timestep consists of 
\begin{equation}
    \label{eq:wagner_reward}
    \MDPrewardvar[\currenttimestep+1] = 1 - \frac{|\liftcoef[,\currenttimestep]|}{\scaleliftcoef}.
\end{equation}
The value 1 assigns a reward for continuing successfully during that timestep within the allowed threshold; the negative part penalizes a lift that is not zero, and this penalty increases linearly as the lift magnitude increases. Even though the positive part of the reward is not required to learn to minimize lift variations, we found that the training improved if we included it. With a maximum episode length of 200 timesteps, the maximum possible episode return is 200, occurring when the lift is zero throughout the episode.

In a full viscous environment, even if it were possible to observe the entire flow state, it would be intractable to train an agent to act on that observation except with some specialized policy architectures. The classical unsteady aerodynamics environment does not suffer from this same limitation because it can be accurately modeled using only two states that represent the wake. It, therefore, allows us to try out and compare the reinforcement learning framework with three types of agents that observe different degrees of the full flow state. All observe the previous lift $\previousliftcoef$ and the current angular states $\AOA[\currenttimestep]$ and $\AOAdot[\currenttimestep]$ but observe different amounts of information about the wake: one that observes no wake information, one that observes pressure information at the airfoil, and one that observes all the system states of the state-space system~\eqref{eq:text_jones_ss_state}, including the wake states (which we will refer to as `full wake'). For a fair comparison, all the observed values, except for $\AOA$ and $\AOAdot$, are subject to the same output delay $\controldelay$. The pressure-observing agent observes the pressure at two points on the plate. This pressure is the sum of the added mass pressure and the circulatory pressure, computed using Eqs.~\eqref{eq:pressure_am}-\eqref{eq:pressure_cir} from Appendix~\ref{app:classical_aero}. It follows from those equations that sensor measurements from two unique positions anywhere along the plate allows the agent to distinguish the two contributions and that any additional pressure observations do not introduce additional independent information. To study the effect of memory on the agent's performance in this POMDP, we will compare two variants of the previous three types of agents: one that only uses the most recent observation and one that retains a memory of the two most recent observations.

\subsection{Training and evaluation}

\begin{figure}
    \centering
    \includegraphics[]{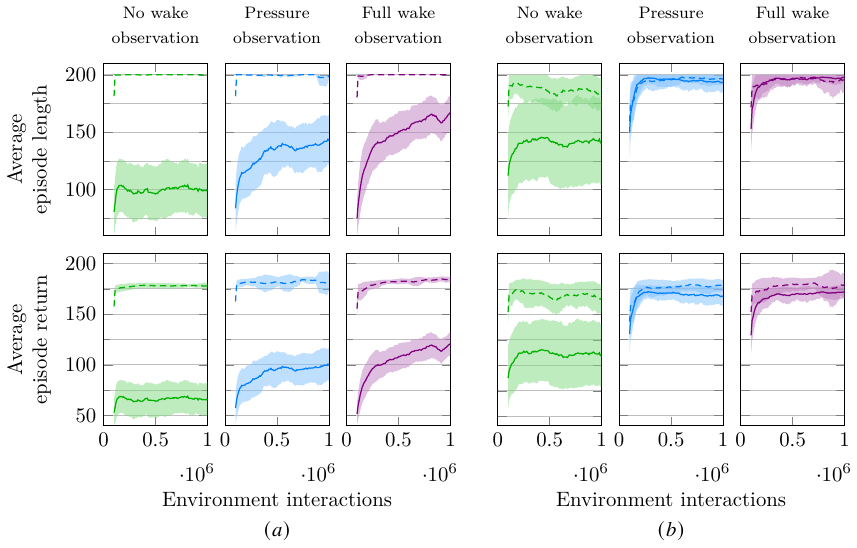}
    \caption{Mean and standard deviation (shade) of the average episode length and return for agents observing no wake information (green), observing pressure (blue), and observing the full wake (violet), with access to the last one (solid lines) or two (dashed lines) observations. Panels $(a)$ and $(b)$ correspond to agents trained and evaluated on episodes with randomized smooth and impulsive $\pivotyaccel$ variations, respectively.}
    \label{fig:wagner_training}
\end{figure}

To mitigate variance stemming from initial random actions and policy exploration, we train each agent variant on both types of disturbance across ten environments initialized with random seeds, resulting in ten policies for each agent. Each policy is trained with one million environment interactions. We evaluate the agents at checkpoints that occur after every 10000 interactions of training. To evaluate an agent, we simulate ten random episodes of one disturbance type with each of its ten policies without applying action noise. We average the 100 $(=10\times 10)$ episode lengths and returns for each evaluation checkpoint and then create a smoothed \emph{training curve} by applying a moving average over ten evaluation checkpoints. These training curves are shown for each possible combination of agent and disturbance type in Figure~\ref{fig:wagner_training} to compare the training progression. 

As anticipated, both the extra wake information and memory help to partially restore the Markovian property and increase the average episode return, with the memory having the largest effect. In this simplified flow problem, the pressure provides sufficient additional information to approximately match the performance of an agent that has knowledge of the entire flow at the previous timestep, especially when the agent has a memory of multiple observations, in which case their performances become indistinguishable for the chosen environment parameters. This confirms that the learning algorithm is able to exploit the pressure information for improved performance, which we will seek to replicate in a viscous environment in the next section, when full-wake observations are unavailable. Additionally, we note that the difference in performance between the agents that do retain memory and those that do not is much smaller for the impulsive disturbances than it is for the smooth disturbances, indicating that there is a fundamental difference for the RL training between both types of disturbances.

\begin{figure}
    \centering
    \includegraphics{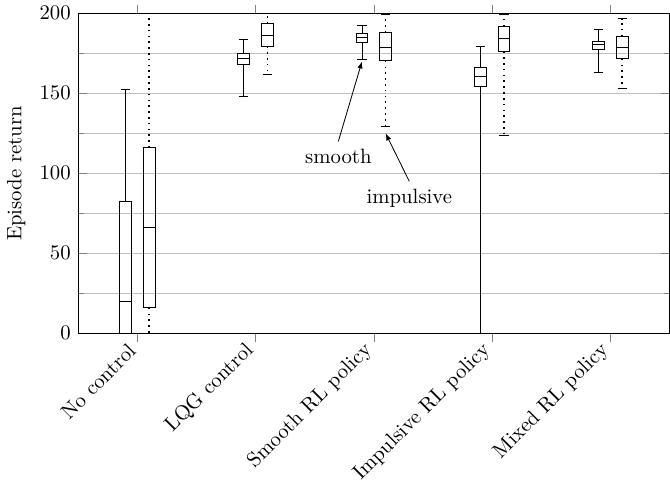} 
    \caption{Comparison of the episode return for different agents for 1000 random episodes each of smooth (solid), and impulsive (dotted) $\pivotyaccel$ variations in the classical unsteady aerodynamics environment. The lower and upper whisker represent the minimum and maximum values of the sampled episodes, respectively.}
    \label{fig:wagner_evaluation}
\end{figure}

As a benchmark for the performance of the RL policies, we tune a discrete linear-quadratic-Gaussian controller for the discrete-time system of~\eqref{eq:text_jones_ss_state}-\eqref{eq:text_jones_ss_output} using output feedback, subject to the same output delay and assuming no measurement noise, resulting in a $z$-domain transfer function
\begin{equation}
    K(z) = \frac{ 1.1836 (z+0.01069) (z^2 - 1.826z + 0.8353)}{(z+1.237) (z-1) (z-0.8592)}.
\end{equation}

\begin{figure}
    \centering
    \includegraphics{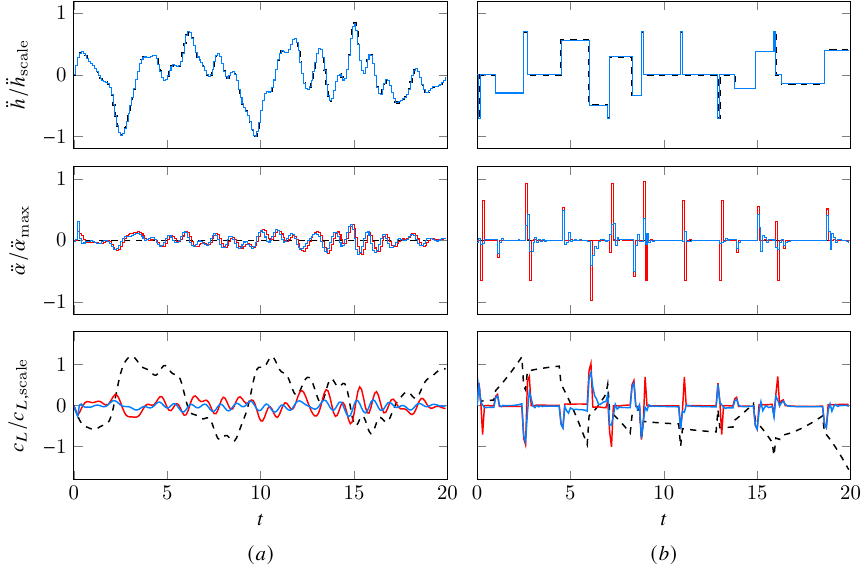} 
    \caption{The vertical acceleration (disturbance), angular acceleration (control input), and lift coefficient for a single episode with $(a)$ smooth disturbances and $(b)$ impulsive disturbances using different controllers: no control (\ref{ls:no_controller_dashed}), LQG control (\ref{ls:p_3}), and the best available RL agent (\ref{ls:myblue}).}
    \label{fig:wagner_episode_stats}
\end{figure}

Figure~\ref{fig:wagner_evaluation} compares the cases with no control, LQG control, and RL control using boxplots that indicate the minimum, first quartile, median, third quartile, and maximum episode returns for a set of episodes. Each boxplot represents the data of 1000 episodes with randomized disturbances of one type. For the RL cases, these 1000 episodes are collected by simulating 100 episodes with each of the ten policies of the agents that observe the full wake with memory. The figure indicates the ability of the RL control to match the performance of this specific LQG controller for impulsive disturbances and exceed it for smooth disturbances. This is also apparent from the plots in Figure~\ref{fig:wagner_episode_stats} of the lift variation during an actual episode with those disturbances. Interestingly, the impulsive episode results in Figure~\ref{fig:wagner_episode_stats} indicate that the RL control learns not to react too strongly to impulsive changes in favor of a longer-term response.

It is important to stress that there are likely to be linear controllers that achieve better and more robust performance than the one devised here for comparison. The main advantage of the RL control here is the ability to treat the plant as a black box. It is also important to acknowledge that the disturbances in this work are functions with random parameters, meaning their values are not stochastically independent from one timestep to the next. Although this should ideally be reflected by additional states in the environment's state vector, we have chosen not to model this aspect in our current work, as it remains internal to the environment. Nevertheless, an effective RL agent could potentially exploit this behavior, as might be the case here if the policy trained on one disturbance type learned to anticipate that disturbance's behavior over the next few timesteps.

Lastly, Figure~\ref{fig:wagner_evaluation} indicates that while the RL agents perform well on the disturbances they were trained on, this behavior does not automatically transfer to the other type of disturbance. In one case, this evaluation led to one of the policies giving unstable results, indicated by the extreme value of its boxplot's lower whisker. This points to the important fact that the nonlinear controller at the heart of the trained deep RL agent is not guaranteed to generalize in the same way that robust linear controllers do. Therefore, care has to be taken when choosing the training data for a controller that should be robust to all kinds of disturbances. In line with this guidance, we trained ten additional policies in an environment with mixed episodes, meaning that the disturbance type for each episode is randomly chosen to be either smooth or impulsive. Their performance on evaluation episodes of one type is slightly lower than the policies trained on exclusively that type of disturbance. However, their worst performance is considerably better than that of the policies trained on one type and evaluated on the other type.

\section{Learning in a viscous flow environment} \label{sec:viscous_environment}

In this example, we apply airfoil pitch control in an incompressible viscous flow environment at $\Reynolds=200$, which poses a greater challenge to the agent due to its high dimensionality and non-linear dynamics. In particular, we study if pressure measurements can improve the agent's performance as it did in a classical unsteady aerodynamics environment. The airfoil is again a flat plate that pitches about the midchord point. Instead of disturbances in the form of $\pivotyaccel$ variations, we focus on local flow disturbances upstream of the airfoil introduced by a localized, transient forcing, resulting in vortex dipoles that move past the airfoil. This environment is illustrated in the bottom panel of Figure~\ref{fig:combined_diagram}.

\subsection{Setup}

The timestep between every agent-environment interaction is reduced to 0.03 convective times and we simulate episodes of at most 100 timesteps. The flow disturbance is a forcing field that varies spatially and temporally in a Gaussian manner and is described in the windtunnel reference frame coordinates as 
\begin{equation}
    \label{eq:forcing}
    \forcefield(x,y,t) = \rho \pivotxvel ^ 2 \chord \frac{(\xforceamplitude, \yforceamplitude)}{\pi^{3/2} \sigma_x \sigma_y \sigma_t} \exp \left( -\frac{\left( x - x_0 \right)^2}{\sigma_x^2} - \frac{\left( y - y_0 \right)^2}{\sigma_y^2} - \frac{\left( t - t_0 \right)^2}{\sigma_{t}^2} \right),
\end{equation}
where $t_0 = 0.5$, $\sigma_t=0.2$, and $x_0 = -0.5$; thus, the forcing is applied at a constant horizontal distance upstream of the airfoil. The other parameters are sampled from uniform continuous distributions as follows. The spatial widths $\sigma_x$ and $\sigma_y$ are chosen from the interval $[0.04,0.1]$ and the vertical position $y_0$ from the interval $[-0.25,0.25]$. Since the temporal and spatial Gaussian widths are relatively small, the forcing appears as pulsed point forces, which is how we will refer to them in the rest of this work. Such a force produces a vortex dipole, which self-propels in the direction in which the force was applied (in addition to being convected by the background flow). The horizontal amplitude $\xforceamplitude$ is sampled from the interval $[0,3]$ and the vertical amplitude $\yforceamplitude$ from the interval $[-3 (y_0 - 0.1)/0.35, -3 (y_0 + 0.1)/0.35]$. These values are chosen so that the vortex dipoles are aimed towards the airfoil and are strong enough to induce a non-linear lift response. We will also investigate the response to two or three of these point force pulses in the same episode, in which case the forcing field is a superposition of multiple instances of~\eqref{eq:forcing}, using independently sampled parameters. To spread the pulses in time, $t_0$ of a new pulse is delayed with respect to that of the previous pulse with a delay uniformly sampled from an interval $[0,1.5]$; the first pulse remains at $t_0 = 0.5$.

The flow simulation is performed using a viscous incompressible flow solver employing an immersed boundary projection method~\citep{eldredge2022} in a body fixed reference frame with a grid cell size $\dx = 3 \chord / \Reynolds$, and a time step size $\dt = 0.4 \dx / \freestreamxvel$ on a domain that spans $[-1.25 \chord, 1.5\chord] \times [-0.5\chord, 0.5\chord]$ with free-space boundary conditions. The flow solver is vorticity-based and utilizes a lattice Green's function for solutions of the Poisson equation, enabling a more compact domain than typical computational aerodynamics studies. The vortex disturbances remain between the upper and lower boundaries and freely convect through the downstream boundary. The flow solver time step in this case equals 0.006 convective times, but we set the agent-environment interaction timestep to 0.03 convective times, chosen to limit the controller bandwidth and create a more challenging control problem.

We again define a scale $\scaleliftcoef = 0.1$ and a maximum control input magnitude $\maxAOAddot = 10$. In addition, we also impose the constraints $|\AOA| < 60\pi/180$ to avoid unrealistic configurations and $|\AOAdot| < 1.8$ to maintain a reasonable viscous flow solver timestep, and we always terminate the episode if these are violated. During training only (and not when evaluating a policy), we terminate an episode if $|\liftcoef|$ exceeds $0.8$.

We modify the reward function from the previous section with two additions. We add a negative reward for high variations of $\AOAddot$ between two timesteps to prevent rapid oscillatory behavior and another negative reward when any of the constraints are violated:
\begin{equation}
    \label{eq:viscous_reward}
    \MDPrewardvar[\currenttimestep+1] = 1 - \frac{|\liftcoef[,\currenttimestep]|}{\scaleliftcoef} - 2 \frac{|\AOAddot[\currenttimestep] - \AOAddot[\currenttimestep - 1]|}{\maxAOAddot} -
    \begin{cases}
        100, \MDPstatevar[\currenttimestep] \not\in \MDPstatespace \\
        0, \text{otherwise}
    \end{cases}
\end{equation}
where $\MDPstatespace$ is the space of valid values of the environment's state, when $\AOA$, $\AOAdot$, and $\liftcoef$ are within their prescribed ranges. With a maximum episode length of 100 timesteps, the maximum possible episode return is 100, again occurring when the lift is zero throughout the episode.

While it would be possible in a simulation to observe the entire vorticity field (and thus the state of the flow) and feed this to a convolutional layer in the policy network, no good results could be obtained with this approach with the TD3 algorithm in our setup. We limit ourselves, therefore, to the following two types of agents: one that observes the current $\AOA[\currenttimestep]$, $\AOAdot[\currenttimestep]$, and the latest $\liftcoef$ computed by the flow solver (with an output delay of one flow solver timestep), and one that additionally observes the latest pressure jump over the plate at seven positions spread equidistantly between 30\% to 70\% chord. Both types of agents are implemented with a memory of the last four observation vectors, stacked together to form the input to their policy. We found that increasing the memory to retain more than four observations did not significantly increase the performance of the controller for the size of the neural network used in this work. Lastly, even though the previous action $\AOAddot[\currenttimestep - 1]$ appears in the reward function, we found that including its value in the observations did not increase the performance.

\subsection{Training and evaluation}

We train the policies of the two types of agents over 500000 interactions on two types of training episodes: one type consisting of a single force pulse disturbance and one consisting of two force pulse disturbances. Each force pulse has new random parameters for every new episode. Each case is run with ten environments initialized with random seeds, resulting in ten policies for each of the four possible combinations of agents and training data. To evaluate the policies and study their generalization to unseen cases, we also use a third type of episode, consisting of three random force pulses, which results in higher disturbed flows with increased nonlinear vortex interactions.

\begin{figure}
    \centering
    \includegraphics{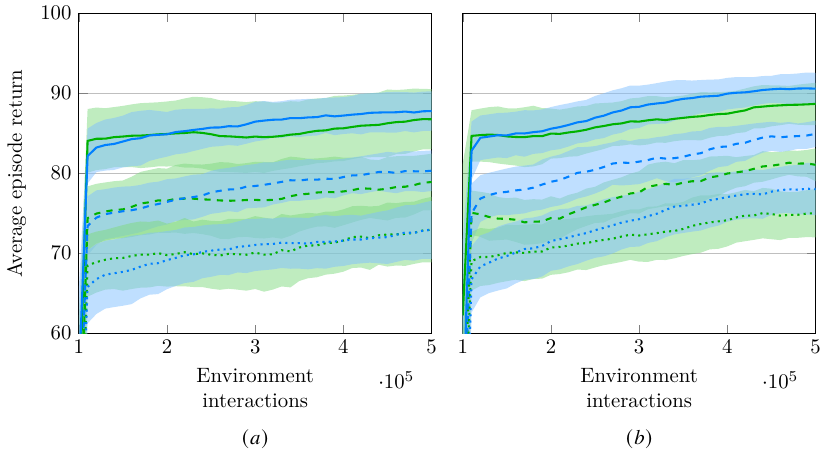} 
    \caption{Mean and standard deviation (shade) of the average episode return for agents observing pressure (blue) and not observing pressure (green), evaluated on episodes consisting of one (solid lines), two (dashed lines), and three (dotted lines) random point force pulses. Panels $(a)$ and $(b)$ correspond to agents trained on episodes with one and two random point force pulses, respectively.}
    \label{fig:viscous_training}
\end{figure}

Figure~\ref{fig:viscous_training} compares the training curves of the different agents, using the same evaluation and averaging procedures as in the classical unsteady aerodynamics environment. The average episode length during training is not shown but rapidly increases to its maximum possible value after approximately 20000 interactions. This indicates that agents quickly learn to avoid the large negative rewards associated with the early episode termination when permissible bounds are violated. After this initial stage, the reward gradually increases as more experience is obtained and processed in the training.

Training an agent with two-pulse episodes instead of one-pulse episodes provides richer training data, including experiences with disturbances that interact with each other. The training curves show that this only has a small effect on the performance of the agents that do not observe pressure. However, it significantly increases the performance of the pressure-based agents, which then, on average, consistently outperform the agents that do not observe pressure. This increased performance also results in a higher average episode return for single-pulse episodes compared to the agents trained exclusively on single-pulse episodes, for the number of interactions in this training.

\begin{figure}
    \centering
    \includegraphics{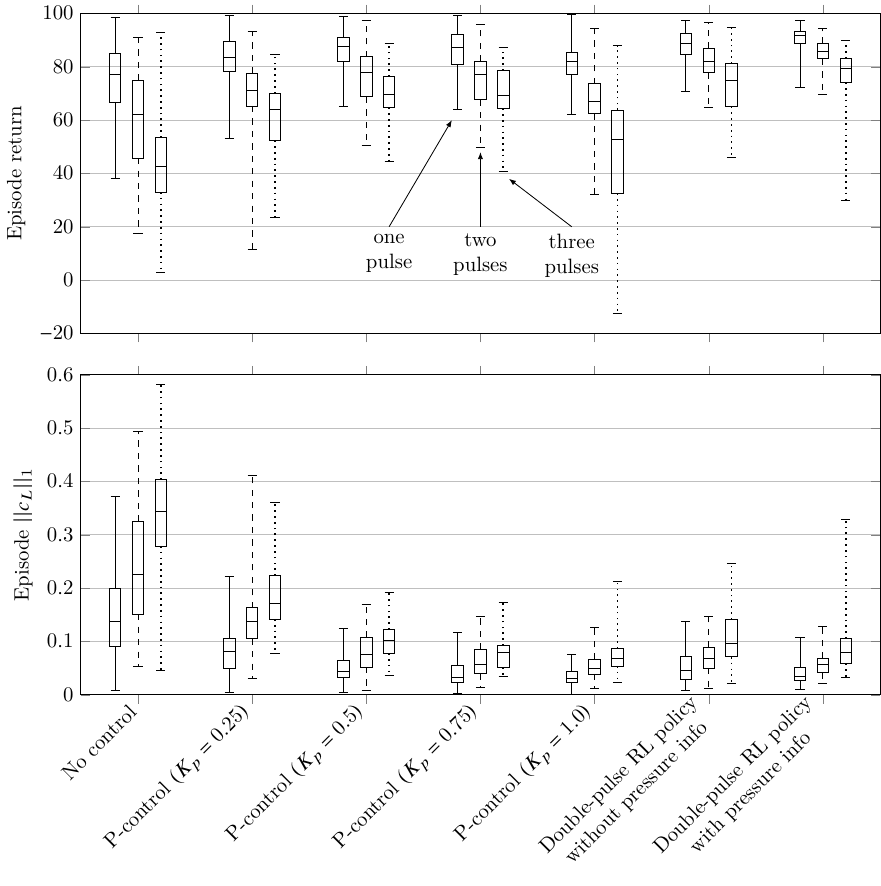} 
    \caption{Comparison of the episode return and lift norm for different agents for 100 random episodes each of one (solid), two (dashed), and three (dotted) point force pulses. The lower and upper whisker represent the minimum and maximum values of the sampled episodes, respectively.}
    \label{fig:viscous_double_aimed_pulse_evaluation}
\end{figure}

\begin{figure}
    \centering
    \includegraphics{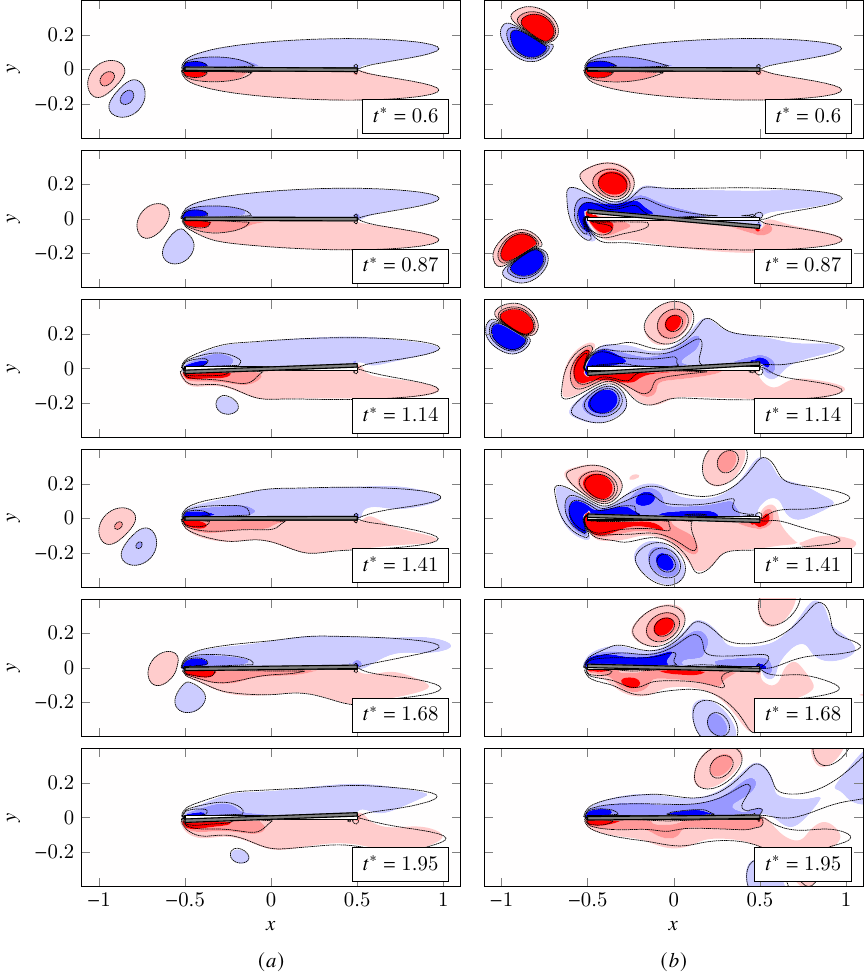} 
    \caption{Snapshots of the vorticity field contours for the best available RL agent (colored contours) and without control (\ref{ls:black_dotted}) at different time instances of $(a)$ a representative episode with two average point force pulses and $(b)$ an extreme episode with three strong force pulses. The uncontrolled airfoil is displayed in white and the controlled airfoil in dark grey. The vorticity contours are drawn in both cases for the values -200, -15, -9, -3, 3, 9, 15, 200.}
    \label{fig:combined_viscous_aimed_point_force_snapshots}
\end{figure}

\begin{figure}
    \centering
    \includegraphics{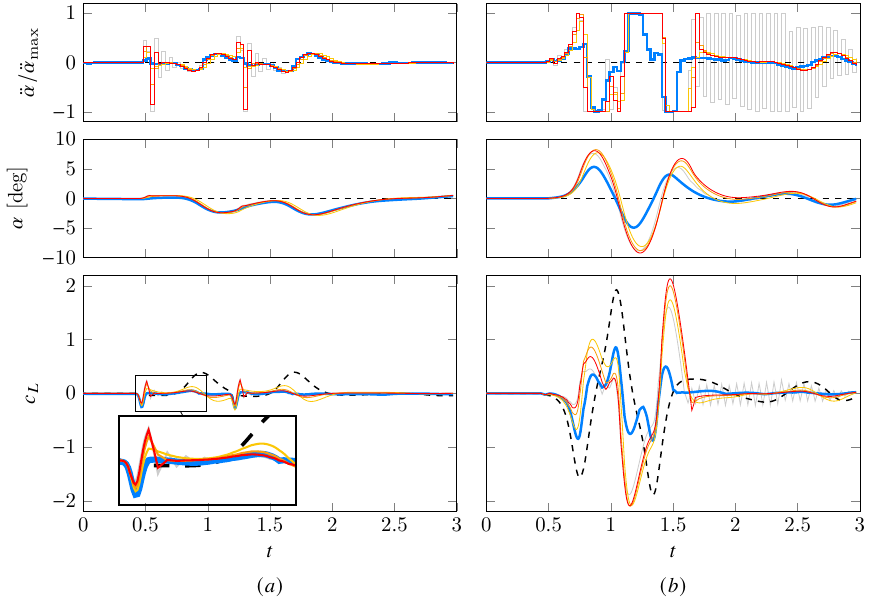} 
    \caption{The angular acceleration, angle of attack, and lift coefficient for $(a)$ a representative episode with two average point force pulses and $(b)$ an extreme episode with three strong force pulses using different controllers: no control (\ref{ls:no_controller_dashed}), proportional control with $\proportionalgain = $ 0.25 (\ref{ls:p_1}), 0.5 (\ref{ls:p_2}), 0.75 (\ref{ls:p_3}), 1.0 (\ref{ls:p_unstable}), and the best available RL agent (\ref{ls:myblue_thick}).}
    \label{fig:viscous_aimed_point_force_episode_stats}
\end{figure}

To evaluate the performance of the agents after the training, we compare them to proportional control ($\AOAddot / \maxAOAddot = - \proportionalgain \liftcoef / \scaleliftcoef$, subject to the same output delay), first using episodes with randomized force pulses and then with two deterministic episodes. The results of the former are presented in Figure~\ref{fig:viscous_double_aimed_pulse_evaluation}, which displays (as boxplots) the distributions of the episode return and the 1-norm of the lift coefficient over 100 episodes of the three kinds of random evaluation episodes. For the RL agents, these 100 episodes are divided equally over their ten policies. For the comparison with deterministic episodes, we define a representative episode, consisting of two pulses, each with $y_0=-0.125$ and all their other parameters set at the average value from their distributions, visualized on the left side of Figure~\ref{fig:combined_viscous_aimed_point_force_snapshots}. We also define an extreme episode with three pulses, visualized on the right side of the figure, which creates a much higher disturbed flow than any of the training episodes and can be used to assess how well an agent generalizes to such a flow. These pulses appear at $t_0$ equal to 0.5, 0.8, and 1.1, and $y_0$ equal to -0.25, 0.25, and -0.25, respectively. Their amplitudes are set to the strongest values from their distributions and their spatial widths to the lowest value of 0.04. Figure~\ref{fig:viscous_aimed_point_force_episode_stats} displays the values of the control input, pitch, and lift coefficient during these two episodes, while Table~\ref{tab:viscous_cases} lists the return, the 1-norms of the lift coefficient and control input, and the average value of the power coefficient $\powercoef=2 \moment \AOAdot / (\rho \pivotxvel ^ 3 c)$, where $\moment$ is the moment about the midchord point.

For the proportional control, Figure~\ref{fig:viscous_double_aimed_pulse_evaluation} indicates that a higher gain results in higher episode returns and lower lift norms. However, once the gain is approximately one or higher, the control input displays rapid oscillations (visible in Figure~\ref{fig:viscous_aimed_point_force_episode_stats}) that are eventually stabilized by the control input bounds. This results in lower episode returns because the reward penalizes fast control input variations. 

For the RL control, Figure~\ref{fig:viscous_double_aimed_pulse_evaluation} and Table~\ref{tab:viscous_cases} illustrate that the episode returns for the random and representative deterministic episodes are on average considerably higher than those achieved without control or with proportional control. However, the lift norm is approximately the same magnitude, indicating that RL can achieve similar lift mitigation performance with less control input variation. On the other hand, the extreme deterministic episode shows the superior performance of the RL control compared to the proportional control in lift mitigation and control expenditure. Figure~\ref{fig:combined_viscous_aimed_point_force_snapshots} shows that the strong vortex dipoles in this episode interact significantly with the leading edge of the plate, the wake of the airfoil, and the previous vortex dipoles. Such strong flow interactions exhibit substantial nonlinear behavior, further compounded by their nonlinear dependence on the pitch variations. This complexity severely compromises the effectiveness of proportional control, leading to inferior performance compared to the uncontrolled scenario. In contrast, the RL control performs markedly better than the uncontrolled case, underscoring its capability of controlling the nonlinear dynamics. For both deterministic episodes, Table~\ref{tab:viscous_cases} shows that the RL control requires less power for better lift mitigation, even though the reward function does not consider energy expenditure.

Surprisingly, the data presented in Table~\ref{tab:viscous_cases} shows that the performance of the agent that does not observe pressure during the extreme episode is better than that of the pressure-observing agent. Similarly, in Figure~\ref{fig:viscous_double_aimed_pulse_evaluation}, the increased range of its boxplot for the three-pulse episodes suggests poorer performance by the pressure-observing agent, despite its higher median performance. This phenomenon may be attributed to a faster convergence during training for the agent that does not observe pressure, as opposed to the pressure-observing agent, whose training may not have fully converged. This discrepancy could primarily affect performance during generalization and extreme cases.

When analyzing these results, one should remember that the RL agents are conditioned through the reward function to find a trade-off between low lift magnitude and small control input variations. This trade-off is further affected by a conservative policy near state thresholds that would result in high negative rewards if exceeded. Designing reward functions with different weights for these contributions (or including different terms) can significantly impact the policy, which we do not explore in this work.

\setlength{\tabcolsep}{5pt}
\begin{table}
    \centering
    \begin{tabular}{c  c *{7}{c}}
        &  & \rotatebox{90}{\thead[l]{No control}} & \rotatebox{90}{\thead[l]{P-control\\($\proportionalgain = 0.25$)}} & \rotatebox{90}{\thead[l]{P-control\\($\proportionalgain = 0.5$)}} & \rotatebox{90}{\thead[l]{P-control\\($\proportionalgain = 0.75$)}} & \rotatebox{90}{\thead[l]{P-control\\($\proportionalgain = 1.0$)}} & \rotatebox{90}{\thead[l]{Double-pulse RL \\ policy without \\ pressure info}} & \rotatebox{90}{\thead[l]{Double-pulse RL \\ policy with \\ pressure info}}

        \bigstrut[b]\\
        \hline
        \bigstrut[t]
         
        & \thead{return} & 58.8  & 70.6 & 76.5 & 73.8 & 63.1 & \num[separate-uncertainty]{79.4 \pm 3.4} & \num[separate-uncertainty]{84.5 \pm 1.3}  \bigstrut \\
        & \thead{$ ||\liftcoef||_1 $} & 0.247 & 0.144 & 0.082 & 0.062 & 0.057 & \num[separate-uncertainty]{0.077 \pm 0.021} & \num[separate-uncertainty]{0.056 \pm 0.008}  \bigstrut \\
        & \thead{$ ||\AOAddot[k]||_1 $} & 0.00 & 3.64 & 4.22 & 4.96 & 6.20 & \num[separate-uncertainty]{3.97 \pm 0.24} & \num[separate-uncertainty]{3.50 \pm 0.27}  \bigstrut \\
        \multirow{-4}[4]{*}{\rotatebox{90}{\thead[c]{Two average\\force pulses}}} & \thead{$ \overline{\powercoef}$ ($\cdot 10^{-3}$)} & 0.00 & 4.15 & 4.11 & 4.26 & 4.43 & \num[separate-uncertainty]{3.56 \pm 0.14} & \num[separate-uncertainty]{3.53 \pm 0.16}

        \bigstrut[b]\\
        \hline
        \bigstrut[t]
         
        & \thead{return} & -77.2 & -90.9 & -82.8 & -79.9 & -209.0 & \num[separate-uncertainty]{0.1 \pm 10.5} & \num[separate-uncertainty]{-26.3 \pm 11.1}  \bigstrut \\
        & \thead{$ ||\liftcoef||_1 $} & 1.063 & 1.032 & 0.965 & 0.925 & 0.787 & \num[separate-uncertainty]{0.489 \pm 0.055} & \num[separate-uncertainty]{0.637 \pm 0.067}  \bigstrut \\
        & \thead{$ ||\AOAddot[k]||_1 $} & 0.00 & 18.65 & 20.60 & 21.34 & 38.27 & \num[separate-uncertainty]{14.42 \pm 1.33} & \num[separate-uncertainty]{16.04 \pm 1.01}  \bigstrut \\
        \multirow{-4}[4]{*}{\rotatebox{90}{\thead[c]{Three strong\\force pulses}}} & \thead{$\overline{\powercoef}$ ($\cdot 10^{-3}$)} & 0.00 & 102.52 & 116.04 & 118.22 & 115.14 & \num[separate-uncertainty]{74.21 \pm 6.34} & \num[separate-uncertainty]{86.71 \pm 6.71}

        \bigstrut[b]\\
        \hline
    \end{tabular}
    \caption{Comparison of different agents for a representative episode with two average point force pulses and an extreme episode with three strong point force pulses.}
    \label{tab:viscous_cases}
\end{table}

\section{Conclusions}

In this work, we studied the use of deep reinforcement learning and the role of partial observability in designing an airfoil pitch controller to minimize lift variations in randomly disturbed flows. In most scenarios, a practical controller would only be able to sense the flow through force and pressure sensors. This information is often not enough to observe the entire state of the flow, which, together with the disturbances, determines the exact lift response of the airfoil. From the perspective of reinforcement learning, the pitch controller is an agent that receives non-Markovian observations from the environment and the control problem can be viewed as a partially-observable Markov decision process (POMDP). Deep reinforcement learning has proven to be a viable approach for approximating an optimal control policy for a given reward function in such conditions. We specifically apply deep RL using the TD3 algorithm to learn pitch control of a flat plate and study agents with and without a memory of earlier observations and with different degrees of environment observations.

We tested our reinforcement learning setup in two types of environments. The first environment modeled the flow and disturbances in a Wagner/Theodorsen setup: the flow is a potential flow with a flat vortex sheet wake and the disturbances are random vertical accelerations of the plate. When analyzing the results in this classical unsteady aerodynamics environment, we confirmed that (1) augmenting the observed information with pressure and wake states or (2) retaining a memory of the two most recent observations both increase the performance of our RL control. This was expected because both aim to restore the Markovian property of the POMDP. We found that our RL control can generally match or exceed the performance of an LQG controller when minimizing the lift variations. Both the RL control and LQG can be further finetuned for specific performance metrics, which we did not consider in this work. Finally, we found that training on smooth disturbances does not guarantee adequate performance on impulsive disturbances and vice versa, even though the underlying problem is linear. This can be explained by the nonlinearity of the RL policy and should be taken into account when designing a robust controller.

The second environment was a viscous flow model where flow disturbances are introduced by pulsed point forces upstream of the airfoil. When analyzing the results in this environment, we found that the RL setup can again take advantage of pressure information in the observations to obtain better policies for lift mitigation. The performance of the RL control improves when two force pulse disturbances instead of one are included in each training episode, in which case the RL control also generalizes better to episodes with three force pulse disturbances. This can be explained by the fact that these two-pulse episodes contain much richer training examples that include the interactions between disturbances. Experience with those interactions is valuable also when acting in situations with more than two disturbances. When comparing the performance with proportional control, we found that the RL control can achieve similar performance with less control input variation for moderate episodes and significantly outperform the proportional control for an episode with very strong disturbances, in which case the flow behavior is strongly nonlinear.

The problem setup in this work can be further explored in multiple ways. Firstly, all of the examples in this work can straightforwardly be extended to the servo problem where the lift has to track a nonzero reference value. Secondly, the two-dimensional viscous simulations in this work are performed at a relatively low Reynolds number. Environments with a higher Reynolds number or three-dimensional flows could become too computationally expensive in which to train a TD3 agent. In these cases, a more sample-efficient algorithm, such as a model-based one, could be a more viable RL approach. Finally, while the exact specification of the reward function has a critical effect on the performance of the RL control, we have not yet explored its connection with the optimality or robustness of the trained control policies.

\section*{Acknowledgments} Support for this work by the National Science Foundation under Award number 2247005 is gratefully acknowledged.

\appendix

\section{Reinforcement learning algorithm details}
\label{app:rl_details}

The reinforcement learning implementation used in this work is the Stable Baselines3~\citep{stable-baselines3} implementation of the twin-delayed deep deterministic policy gradient algorithm (TD3)~\citep{fujimoto2018}. The TD3 algorithm utilizes six neural networks: one policy, or actor, networks, two state-action value (Q-function) approximation, or critic, network, and their target networks. These networks all have the same feedforward architecture: an input layer whose size depends on the observations, two fully-connected hidden layers with 400 and 300 rectified linear units (ReLU), respectively, and an output layer representing the action in the case of an actor network or the state-action value in case of a critic network. The target networks are copies of the actor and critic networks with weights that lag behind their counterparts' weights through polyak averaging as $\weightscritictarget = 0.005 \weightscritictarget + 0.995 \weightscritic$. Their purpose is to provide a fixed target when updating their counterpart networks, which prevents training instabilities due to the recursiveness of the underlying Bellman equation.

The training algorithm stores agent-environment interactions in a standard first-in-first-out replay buffer, which can store up to 1000000 interactions. Each interaction consists of the current observation, the chosen action, the resulting subsequent observation, and the corresponding reward. Once 10000 interactions have been generated with a random policy, the training starts and the algorithm switches repeatedly between performing a training step and simulating a new episode with the latest trained policy to further update the replay buffer. During each training step, the networks are updated multiple times using stochastic gradient descent combined with the Adam optimization algorithm using a learning rate of 0.001. Each stochastic gradient descent update uses a batch of 256 randomly sampled interactions from the replay buffer. The critic networks and their targets are updated as many times as there were steps in the last generated episode, and the actor network and its target are updated during every second update of the critic networks using the same batch of interactions. The training uses a discount factor of 0.98 for future rewards and applies target policy smoothing using zero-mean Gaussian noise with a standard deviation of 0.1. This noise is added to the action generated by the target policy network and regularizes the Q-value approximations of the critic networks. Lastly, the algorithm adds zero-mean Gaussian noise with a standard deviation of 0.1 to the actions of the policy when collecting training data to promote the exploration of the state and action spaces.

\section{Classical unsteady aerodynamics model}
\label{app:classical_aero}

\citet{wagner1925} and \citet{theodorsen1935} analyzed the aerodynamic response of an infinitesimally thin flat plate airfoil to specific flow disturbances under the following assumptions. The flat plate has chord length $\chord$ and is at a small angle of attack $\AOA$, where $|\AOA| \ll 1$. Its main motion is a constant velocity $\pivotxvel$ in the negative $x$-direction. Any other motions are perturbations of the angle of attack with an angular velocity $\AOAdot$ applied at a pivot point at a coordinate $(\pivotdist,0)$ in the body-fixed reference frame positioned at the center of the airfoil with its $\coordbody{x}$-axis along the chord and the vertical motion $\pivotyvel$ of that pivot point. These perturbations are considered small compared to the forward velocity of the plate, i.e., $|\AOAdot| \ll \pivotxvel / \chord$ and $|\pivotyvel| \ll \pivotxvel$. The wake is assumed to lie along the $\coordbody{x}$-axis of the body, and the wake vorticity convects with the velocity $\pivotxvel$ away from the airfoil.

Classical unsteady aerodynamics theory~\citep{vonkarman1938} provides an expression for the bound vortex sheet strength when the Kutta condition is applied at the trailing edge under the previously mentioned assumptions:
\begin{equation}
    \label{eq:classic_kutta_condition}
    \int_{\chord/2}^{\infty} \frac{(\coordbody{x} + \chord/2)^{1/2}}{(\coordbody{x} - \chord/2)^{1/2}} \vortsheetstrengthwake (\coordbody{x}) \mathrm{d}\coordbody{x} = -\cirqs
\end{equation}
where the quasi-steady circulation $\cirqs \coloneqq - \pi \chord \pivotxvel \AOAeff$ depends on the effective angle of attack
\begin{equation}
    \AOAeff \coloneqq -\frac{\pivotyvel}{\pivotxvel} + \AOA + \left( \frac{\chord}{4} - \pivotdist \right) \frac{\AOAdot}{\pivotxvel},
\end{equation}

Assuming that the wake convects with the velocity $\pivotxvel$ and equation~\eqref{eq:classic_kutta_condition} has to be satisfied at any instant of time, the wake strength can be determined.
Furthermore, the classical theory also provides a general expression for the lift on the plate:
\begin{equation}
    \label{eq:classic_aero_lift}
    \lift = {\underbrace{\vphantom{\int_{\chord/2}^{\infty}}-\frac{\pi}{4}\rho \chord^2 \left(\pivotyaccel + \pivotdist \AOAddot - \pivotxvel \AOAdot \right)}_{\text{added-mass lift $\liftam$}}} \quad {\underbrace{\vphantom{\int_{\chord/2}^{\infty}} - \rho \pivotxvel \cirqs}_{\substack{\text{quasi-steady} \\ \text{lift $\liftqs$}}}} -  {\underbrace{\frac{1}{2}\rho \pivotxvel \chord \int_{\chord/2}^{\infty} \frac{\vortsheetstrengthwake (\bodyxwake)}{(\bodyxwake^2 - \chord^2/4)^{1/2}} \mathrm{d}\bodyxwake}_{\text{wake lift $\liftwake$}}}.
\end{equation}
The last two terms (the quasi-steady and wake lift) together form the circulatory lift, or $\liftcir = \liftqs + \liftwake$.

The distributions of the pressure jump across the plate corresponding to the added-mass and circulatory lift were first derived by \citet{neumark1952} and are for rigid-body motion given by
\begin{gather}
    \label{eq:pressure_am}
    \diff{\pressuream}(\bodyxplate) = -2 \rho \left(\pivotyaccel + \pivotdist \AOAddot - \pivotxvel \AOAdot \right) \left( \chord^2 / 4 - \bodyxplate^2 \right)^{1/2} \\
    \label{eq:pressure_cir}
    \diff{\pressurecir}(\bodyxplate) = \left( -2 \rho \pivotxvel^2 \AOAeff + \frac{\rho \pivotxvel}{\pi} \int_{\chord/2}^{\infty}\frac{\vortsheetstrengthwake (\bodyxwake)}{(\bodyxwake^2 - \chord^2/4)^{1/2}} \mathrm{d}\bodyxwake\right) \frac{ \left( \chord/2 - \bodyxplate \right)^{1/2}}{\left( \chord/2 + \bodyxplate \right)^{1/2}}.
\end{gather}
The full pressure is then $\diff{\pressure} = \diff{\pressuream} + \diff{\pressurecir}$. Note that the first factor in parentheses in Eq.~\eqref{eq:pressure_cir} is equal to $-2 \liftcir / \pi \chord$ and thus the circulatory pressure at any point along the airfoil linearly scales with the circulatory lift.

For an arbitrary, smoothly-varying $\AOAeff$, a more practical expression can be obtained for the circulatory lift if we formulate it as the convolution of the rate of change of $\cirqs$ with the non-dimensional lift response to a step change in $\cirqs$, known as Wagner's function $\wagfunc$~\citep{wagner1925}, plus the contribution of the initial value of $\cirqs$ times Wagner's function evaluated at the current time:
\begin{equation}
    \label{eq:wagner}
    \lift = \liftam - \rho \pivotxvel \cirqs(0) \wagfunc(\tconv) - \rho \pivotxvel \int_0^{\tconv} \cirqsdot (\tauconv) \wagfunc(\tconv - \tauconv) \mathrm{d} \tauconv.
\end{equation}

There is no analytical form of Wagner's function available, but several approximations exist, including one proposed by \citet{jones1938} that is expressed as
\begin{equation}
    \label{eq:jones_approx}
    \Phi(\tconv) \approx 1 - 0.165 \mathrm{e}^{-0.091\tconv} - 0.335 \mathrm{e}^{-0.6\tconv}.
\end{equation}

The circulatory lift in~\eqref{eq:wagner} can be formulated in the Laplace domain as a multiplication of a transfer function with the quasi-steady lift:
\begin{align}
    \label{eq:wagner_laplace}
    \laptransf\{\liftcir\} &= - \rho \pivotxvel \cirqs(0) \laptransf\{\wagfunc\} - \rho \pivotxvel \laptransf\{\wagfunc\} \laptransf\{\cirqsdot\} \\
    &= - \rho \pivotxvel s \laptransf\{\wagfunc\} \laptransf\{\cirqs\} \\
    &= s \laptransf\{\wagfunc\} \laptransf\{\liftqs\}
\end{align}
where $\lapvar$ is the Laplace variable. The transfer function $\theofunc(\lapvar) = \lapvar \laptransf\{\wagfunc\}$ based on the Wagner function approximation~\eqref{eq:jones_approx} was also derived by \citet{jones1938} and can be expressed as
\begin{align}
    \label{eq:jones_laplace}
    \theofunc(\lapvar) &= \lapvar \int_0^\infty \wagfunc (\tauconv) e^{-\lapvar\tauconv}\mathrm{d}\tauconv \\
    &\approx \frac{0.5 \lapvar^2 + 0.5616 \lapvar + 0.0546}{\lapvar^2 + 0.691 \lapvar + 0.0546}.
\end{align}
Furthermore, \citet{jones1938} noted, like \citet{garrick1938} before him, that $\theofunc$ evaluated on the imaginary axis, i.e. its steady-state behavior for a purely oscillatory, quasi-steady circulation, is equivalent to Theodorsen's function~\citep{theodorsen1935}.

To design a controller or RL policy based on (partial) state feedback, we seek the state-space representation of~\eqref{eq:jones_laplace} and express the lift in terms in terms of the lift coefficient $\liftcoef = 2 \lift / \left( \rho \pivotxvel^2 \chord \right)$. The controllable canonical state-space representation for $\theofunc(s)$ is
\begin{gather}
    \begin{bmatrix}
        \dot{\theostate}_1 \\
        \dot{\theostate}_2
    \end{bmatrix}
    = 
    \underbrace{
    \begin{bmatrix}
        -0.691 & -0.0546 \\
        1 & 0
    \end{bmatrix}
    \vphantom{
    \begin{bmatrix}
        \theostate[1] \\
        \theostate[2]
    \end{bmatrix}
    }
    }_{\textstyle \sstheoA \mathstrut}
    \begin{bmatrix}
        \theostate[1] \\
        \theostate[2]
    \end{bmatrix}
    +
    \underbrace{
    \begin{bmatrix}
        1 \\
        0
    \end{bmatrix}
    }_{\textstyle \sstheoB \mathstrut}
    \liftcoefqs
    \\
    \liftcoefcir = 
    \underbrace{
    \begin{bmatrix}
        0.2161 & 0.0273
    \end{bmatrix}
    \vphantom{
    \begin{bmatrix}
        \theostate[1] \\
        \theostate[2]
    \end{bmatrix}
    }
    }_{\textstyle \sstheoC \mathstrut}
    \begin{bmatrix}
        \theostate[1] \\
        \theostate[2]
    \end{bmatrix}
    +
    \underbrace{
    \begin{bmatrix}
        0.5
    \end{bmatrix}
    \vphantom{
    \begin{bmatrix}
        \theostate[1] \\
        \theostate[2]
    \end{bmatrix}
    }
    }_{\textstyle \sstheoD \mathstrut}
    \liftcoefqs.
\end{gather}

\begin{figure}
    \centering
    \includegraphics{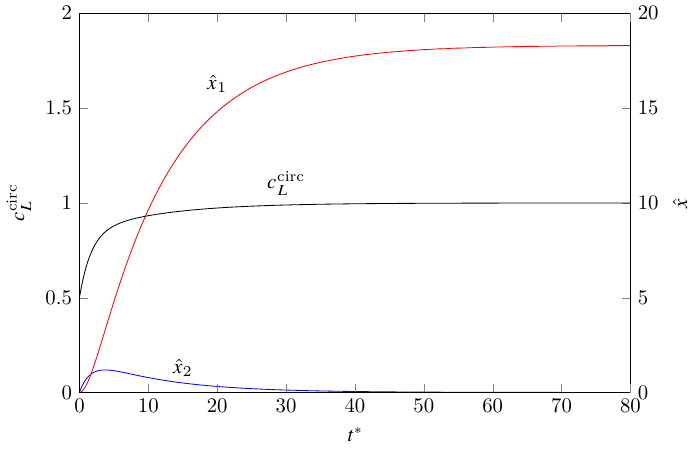} 
    \caption{Unit step response of the circulatory lift coefficient and the states of the controllable canonical state-space representation for the Wagner transfer function by \citet{jones1938}.}
    \label{fig:wagner_step_response}
\end{figure}

The response of this system to a unit step increase of $\liftcoefqs$ is shown in Figure~\ref{fig:wagner_step_response}.

Following \citet{brunton2013}, we can use \eqref{eq:jones_laplace} to create a multi-input, single-output transfer function for the complete lift expression by applying the Laplace transform to~\eqref{eq:wagner} for each input:
\begin{align}
    \label{eq:jones_tf}
    \TFwagner(\lapvar)
    &= 
    \begin{bmatrix}
        \TFwagnerpivotyaccel(\lapvar) & \TFwagnerAOAddot(\lapvar)
    \end{bmatrix} \\
    &=
    \begin{bmatrix}
        \displaystyle -\frac{\pi \chord}{2 \pivotxvel^2} - \frac{2 \pi}{\lapvar \pivotxvel} \theofunc(\lapvar) & \displaystyle \frac{\pi \chord}{2 \pivotxvel} \left( \frac{1}{s} + \frac{\pivotdist}{\pivotxvel} \right) + 2 \pi \left( \frac{1}{s^2} + \frac{1}{\lapvar \pivotxvel} \left( \frac{\chord}{4} - \pivotdist \right)\right) \theofunc(\lapvar)
    \end{bmatrix},
\end{align}
where $\TFwagnerpivotyaccel$ and $\TFwagnerAOAddot$ are the transfer functions that give the reponse of the output $\liftcoef$ to the input signals $\pivotyaccel$ and $\AOAddot$, respectively. We can also cast this result also into a minimal state-space realization:
\begin{gather}
    \label{eq:jones_ss_state}
    \begin{bmatrix}
        \AOAeffdot \\
        \AOAddot \\
        \theoboldstatedot
    \end{bmatrix}
    =
    \underbrace{
    \begin{bmatrix}
        0 & 1 & \vect{0} \\
        0 & 0 & \vect{0} \\
        2 \pi \sstheoB & \vect{0} & \sstheoA
    \end{bmatrix}
    }_{\textstyle \ssA \mathstrut}
    \begin{bmatrix}
        \AOAeff \\
        \AOAdot \\
        \theoboldstate
    \end{bmatrix}
    +
    \underbrace{
    \begin{bmatrix}
        -\frac{1}{\pivotxvel} & \frac{1}{\pivotxvel} \left( \frac{\chord}{4} - \pivotdist \right) \\
        0 & 1 \\
        \vect{0} & \vect{0}
    \end{bmatrix}
    }_{\textstyle \ssB \mathstrut}
    \begin{bmatrix}
        \pivotyaccel \\
        \AOAddot 
    \end{bmatrix}
    \\
    \label{eq:jones_ss_output}
    \liftcoef
    =
    \underbrace{
    \begin{bmatrix}
        2 \pi \sstheoD & \frac{\pi \chord}{2 \pivotxvel} & \sstheoC
    \end{bmatrix}
    \vphantom{\begin{bmatrix}
        \AOAeff \\
        \AOAdot \\
        \theoboldstate
    \end{bmatrix}
    }
    }_{\textstyle \ssC \mathstrut}
    \begin{bmatrix}
        \AOAeff \\
        \AOAdot \\
        \theoboldstate
    \end{bmatrix}
    +
    \underbrace{
    \begin{bmatrix}
        -\frac{\pi \chord}{2 \pivotxvel^2} & -\frac{\pi \chord \pivotdist}{2 \pivotxvel^2}
    \end{bmatrix}
    \vphantom{\begin{bmatrix}
        \AOAeff \\
        \AOAdot \\
        \theoboldstate
    \end{bmatrix}
    }
    }_{\textstyle \ssD \mathstrut}
    \begin{bmatrix}
        \pivotyaccel \\
        \AOAddot 
    \end{bmatrix}
\end{gather}

\bibliographystyle{unsrtnat}
\bibliography{main}

\scalebox{0}{\begin{tikzpicture}
\begin{axis}[hide axis,xmin=1,xmax=2,ymin=1,ymax=2]
    \addplot[myblue,line width=0.5,forget plot](0,0); \label{ls:myblue}
    \addplot[myblue,line width=1.5,solid,forget plot](0,0); \label{ls:myblue_thick}
    \addplot[no_wake_info,line width=1,forget plot](0,0); \label{ls:no_wake_info}
    \addplot[no_controller,line width=1,dashed,forget plot](0,0); \label{ls:no_controller_dashed}
    \addplot[full_wake_info,line width=1,forget plot](0,0); \label{ls:full_wake_info}
    \addplot[pressure_info,line width=1,forget plot](0,0); \label{ls:pressure}
    \addplot[proportional_controller_1,line width=0.5,forget plot](0,0); \label{ls:p_1}
    \addplot[proportional_controller_2,line width=0.5,forget plot](0,0); \label{ls:p_2}
    \addplot[proportional_controller_3,line width=0.5,forget plot](0,0); \label{ls:p_3}
    \addplot[unstable_proportional_controller,line width=1,forget plot](0,0); \label{ls:p_unstable}
    \addplot[only marks, mark=x, mark options={color=red, scale=1.5},forget plot](0,0); \label{ls:control_decision}
    \addplot[only marks, mark=o, mark options={color=red, scale=1.5, line width=0.7},forget plot](0,0); \label{ls:lift_measurement}
    \addplot[only marks, mark=square, mark options={color=blue, scale=1.5, line width=0.7},forget plot](0,0); \label{ls:angular_measurement}
    \addplot[black,line width=1,dotted,forget plot](0,0); \label{ls:black_dotted}
\end{axis}
\end{tikzpicture}}

\end{document}

%% file: aeromacros.tex
\newcommand{\vect}[1]{\mathbf{#1}}
\newcommand{\matr}[1]{\mathbf{#1}}

\newcommand{\coordbody}[1]{\tilde{#1}}
\newcommand{\coordinertial}[1]{{#1}}
\newcommand{\tconv}{t}
\newcommand{\tauconv}{\tau}
\newcommand{\lapvar}{s}

\newcommand{\plate}{\mathcal{P}}
\newcommand{\wake}{\mathcal{S}}

\newcommand{\bodyxwake}{\coordbody{x}_\wake}
\newcommand{\bodyxplate}{\coordbody{x}_\plate}

\newcommand{\currenttimestep}{k}

\newcommand{\MDPactionvar}[1][]{A_{#1}}

\newcommand{\MDPstatevar}[1][]{S_{#1}}
\newcommand{\MDPstatespace}{\mathcal{S}}

\newcommand{\MDPrewardvar}[1][]{R_{#1}}

\newcommand{\POMDPobservationvar}[1][]{O_{#1}}

\newcommand{\weightscritic}{\phi}
\newcommand{\weightscritictarget}{\phi_{\mathrm{target}}}

\newcommand{\laptransf}{\mathcal{L}}

\newcommand{\chord}{c}

\newcommand{\Reynolds}{\mathrm{Re}}

\newcommand{\quasisteady}{\mathrm{qs}}
\newcommand{\circulatory}{\mathrm{circ}}
\newcommand{\addedmass}{\mathrm{am}}
\newcommand{\lift}[1][]{f_{y#1}}
\newcommand{\moment}[1][]{m_{#1}}

\newcommand{\powercoef}[1][]{c_{P#1}}
\newcommand{\liftcoef}[1][]{c_{L#1}}
\newcommand{\previousliftcoef}{\liftcoef[,\currenttimestep-1]}

\newcommand{\liftam}{\lift^{\addedmass}}

\newcommand{\liftcir}{\lift^{\circulatory}}
\newcommand{\liftcoefcir}{\liftcoef^{\circulatory}}
\newcommand{\liftqs}{\lift^{\quasisteady}}
\newcommand{\liftcoefqs}{\liftcoef^{\quasisteady}}
\newcommand{\liftwake}{\lift^{\wake}}

\newcommand{\pressure}[1][]{p_{#1}}

\newcommand{\pressuream}[1][]{\pressure[#1]^{\addedmass}}
\newcommand{\pressurecir}[1][]{\pressure[#1]^{\circulatory}}
\newcommand{\diff}[1]{[#1]^{+}_{-}}

\newcommand{\AOA}[1][]{\alpha_{#1}}
\newcommand{\AOAdot}[1][]{\dot{\AOA}_{#1}}
\newcommand{\AOAddot}[1][]{\ddot{\AOA}_{#1}}
\newcommand{\AOAeff}[1][]{\AOA[\mathrm{eff}#1]}
\newcommand{\AOAeffdot}[1][]{\AOAdot[\mathrm{eff}#1]}

\newcommand{\pivotdist}{d}

\newcommand{\pivotxvel}{U}
\newcommand{\pivotyvel}{\dot{h}}

\newcommand{\pivotyaccel}[1][]{\ddot{h}_{#1}}

\newcommand{\freestreamxvel}[1][]{U_{\infty#1}}

\newcommand{\forcefield}[1][]{\vect{F}_{#1}}

\newcommand{\xforceamplitude}[1][]{{D}_{x#1}}
\newcommand{\yforceamplitude}[1][]{{D}_{y#1}}

\newcommand{\cir}{\Gamma}
\newcommand{\cirqs}{\cir^\quasisteady}
\newcommand{\cirqsdot}{\dot{\cir}^\quasisteady}
\newcommand{\vortsheetstrength}{\gamma}
\newcommand{\vortsheetstrengthwake}{\vortsheetstrength_\wake}

\newcommand{\wagfunc}{\Phi}

\newcommand{\theofunc}{C}
\newcommand{\theovar}[1]{\hat{#1}}

\newcommand{\ssA}{\matr{A}}
\newcommand{\ssB}{\matr{B}}
\newcommand{\ssC}{\matr{C}}
\newcommand{\ssD}{\matr{D}}

\newcommand{\theostate}[1][]{\theovar{x}_{#1}}
\newcommand{\theoboldstate}[1][]{\theovar{\vect{x}}_{#1}}
\newcommand{\theoboldstatedot}[1][]{\dot{\theovar{\vect{x}}}_{#1}}
\newcommand{\sstheoA}{\theovar{\matr{A}}}
\newcommand{\sstheoB}{\theovar{\matr{B}}}
\newcommand{\sstheoC}{\theovar{\matr{C}}}
\newcommand{\sstheoD}{\theovar{\matr{D}}}

\newcommand{\TFgeneralcontroller}{\vect{K}}
\newcommand{\TFwagner}{\vect{G}}
\newcommand{\TFwagnerAOAddot}{G_{\AOAddot}}
\newcommand{\TFwagnerpivotyaccel}{G_{\pivotyaccel}}

\newcommand{\measurement}[1][]{\vect{z}_{#1}}
\newcommand{\proportionalgain}{K_p}
\newcommand{\controldelay}{\tau}

\newcommand{\maxAOAddot}{\AOAddot[\mathrm{max}]}

\newcommand{\scalepivotyaccel}{\pivotyaccel[\mathrm{scale}]}
\newcommand{\scaleliftcoef}{\liftcoef[,\mathrm{scale}]}

\newcommand{\dx}{\Delta x}
\newcommand{\dt}{\Delta t}

%% file: mycolors.tex
\usepackage{xcolor}

\colorlet{mygreen}{green!70!black}
\colorlet{myblue}{blue!50!-red}
\colorlet{myred}{red}
\colorlet{myyellow}{yellow!80!black}
\definecolor{mygrey}{gray}{0.65}
\definecolor{juliaibpmgreen}{RGB}{157,179,124}

\colorlet{full_wake_info}{violet}
\colorlet{pressure_info}{myblue}
\colorlet{no_wake_info}{mygreen}
\colorlet{no_controller}{black}
\colorlet{proportional_controller_1}{yellow!80!red}
\colorlet{proportional_controller_2}{yellow!60!red}
\colorlet{proportional_controller_3}{yellow!0!red}
\colorlet{unstable_proportional_controller}{white!80!black}

%% file: main.bbl
\begin{thebibliography}{40}
\providecommand{\natexlab}[1]{#1}
\providecommand{\url}[1]{\texttt{#1}}
\expandafter\ifx\csname urlstyle\endcsname\relax
  \providecommand{\doi}[1]{doi: #1}\else
  \providecommand{\doi}{doi: \begingroup \urlstyle{rm}\Url}\fi

\bibitem[An et~al.(2016)An, Williams, Eldredge, and Colonius]{an2016}
X.~An, D.~R. Williams, J.~D. Eldredge, and T.~Colonius.
\newblock Modeling dynamic lift response to actuation.
\newblock In \emph{54th AIAA Aerospace Sciences Meeting}, 2016.
\newblock \doi{https://doi.org/10.2514/6.2016-0058}.

\bibitem[Jones(2020)]{jones2020}
A.~R. Jones.
\newblock Gust encounters of rigid wings: Taming the parameter space.
\newblock \emph{Physical Review Fluids}, 5\penalty0 (11):\penalty0 110513,
  2020.
\newblock URL \url{https://doi.org/10.1103/PhysRevFluids.5.11051}.

\bibitem[Skogestad and Postlethwaite(2005)]{skogestad2005}
S.~Skogestad and I.~Postlethwaite.
\newblock \emph{Multivariable feedback control: Analysis and Design}.
\newblock John Wiley, Hoboken, US-NJ, 2005.

\bibitem[Ahuja and Rowley(2010)]{ahuja2010}
S.~Ahuja and C.~W. Rowley.
\newblock Feedback control of unstable steady states of flow past a flat plate
  using reduced-order estimators.
\newblock \emph{Journal of fluid mechanics}, 645:\penalty0 447--478, 2010.
\newblock URL \url{http://dx.doi.org/10.1017/S0022112009992655}.

\bibitem[Sedky et~al.(2022)Sedky, Gementzopoulos, Andreu-Angulo, Lagor, and
  Jones]{sedky2022}
G.~Sedky, A.~Gementzopoulos, I.~Andreu-Angulo, F.~D. Lagor, and A.~R. Jones.
\newblock Physics of gust response mitigation in open-loop pitching manoeuvres.
\newblock \emph{Journal of Fluid Mechanics}, 944:\penalty0 A38, 2022.
\newblock URL \url{https://doi.org/10.1017/jfm.2022.509}.

\bibitem[Brunton and Rowley(2013)]{brunton2013}
S.~L. Brunton and C.~W. Rowley.
\newblock Empirical state-space representations for theodorsen's lift model.
\newblock \emph{Journal of Fluids and Structures}, 38:\penalty0 174--186, 2013.
\newblock ISSN 0889-9746.
\newblock URL \url{https://doi.org/10.1016/j.jfluidstructs.2012.10.005}.

\bibitem[Brunton et~al.(2014)Brunton, Dawson, and Rowley]{brunton2014}
S.~L. Brunton, S.~T.~M. Dawson, and C.~W. Rowley.
\newblock State-space model identification and feedback control of unsteady
  aerodynamic forces.
\newblock \emph{Journal of Fluids and Structures}, 50:\penalty0 253--270, 2014.
\newblock URL \url{https://doi.org/10.1016/j.jfluidstructs.2014.06.026}.

\bibitem[Sedky et~al.(2020)Sedky, Lagor, and Jones]{sedky2020}
G.~Sedky, F.~D. Lagor, and A.~R. Jones.
\newblock Unsteady aerodynamics of lift regulation during a transverse gust
  encounter.
\newblock \emph{Physical Review Fluids}, 5\penalty0 (7):\penalty0 074701, 2020.

\bibitem[Kerstens et~al.(2011)Kerstens, Pfeiffer, Williams, King, and
  Colonius]{kerstens2011}
W.~Kerstens, J.~Pfeiffer, D.~R. Williams, R.~King, and T.~Colonius.
\newblock Closed-loop control of lift for longitudinal gust suppression at low
  reynolds numbers.
\newblock \emph{AIAA Journal}, 49\penalty0 (8):\penalty0 1721--1728, 2011.
\newblock URL \url{https://doi.org/10.2514/1.J050954}.

\bibitem[Bieker et~al.(2020)Bieker, Peitz, Brunton, Kutz, and
  Dellnitz]{bieker2020}
K.~Bieker, S.~Peitz, S.~L. Brunton, J.~N. Kutz, and M.~Dellnitz.
\newblock Deep model predictive flow control with limited sensor data and
  online learning.
\newblock \emph{Theoretical and computational fluid dynamics}, 34:\penalty0
  577--591, 2020.

\bibitem[Xu et~al.(2023)Xu, Gementzopoulos, Sedky, Jones, and Lagor]{xu2023}
X.~Xu, A.~Gementzopoulos, G.~Sedky, A.~R. Jones, and F.~D. Lagor.
\newblock Design of optimal wing maneuvers in a transverse gust encounter
  through iterated simulation or experiment.
\newblock \emph{Theoretical and Computational Fluid Dynamics}, pages 1--20,
  2023.

\bibitem[Arulkumaran et~al.(2017)Arulkumaran, Deisenroth, Brundage, and
  Bharath]{arulkumaran2017}
K.~Arulkumaran, M.~P. Deisenroth, M.~Brundage, and A.~A. Bharath.
\newblock Deep reinforcement learning: A brief survey.
\newblock \emph{IEEE Signal Processing Magazine}, 34\penalty0 (6):\penalty0
  26--38, 2017.
\newblock URL \url{https://doi.org/10.1109/MSP.2017.2743240}.

\bibitem[Sutton and Barto(2018)]{sutton2018}
R.~S. Sutton and A.~G. Barto.
\newblock \emph{Reinforcement learning: An introduction}.
\newblock MIT press, 2018.

\bibitem[Kurniawati(2022)]{kurniawati2022}
H.~Kurniawati.
\newblock Partially observable markov decision processes and robotics.
\newblock \emph{Annual Review of Control, Robotics, and Autonomous Systems},
  5\penalty0 (1):\penalty0 253--277, 2022.
\newblock URL \url{https://doi.org/10.1146/annurev-control-042920-092451}.

\bibitem[Verma et~al.(2018)Verma, Novati, and Koumoutsakos]{verma2018}
S.~Verma, G.~Novati, and P.~Koumoutsakos.
\newblock Efficient collective swimming by harnessing vortices through deep
  reinforcement learning.
\newblock \emph{Proceedings of the National Academy of Sciences}, 115\penalty0
  (23):\penalty0 5849--5854, 2018.
\newblock URL \url{https://doi.org/10.1073/pnas.1800923115}.

\bibitem[Novati et~al.(2019)Novati, Mahadevan, and Koumoutsakos]{novati2019}
G.~Novati, L.~Mahadevan, and P.~Koumoutsakos.
\newblock Controlled gliding and perching through deep-reinforcement-learning.
\newblock \emph{Physical Review Fluids}, 4\penalty0 (9):\penalty0 093902, 2019.
\newblock URL \url{https://doi.org/10.1103/PhysRevFluids.4.093902}.

\bibitem[Gunnarson et~al.(2021)Gunnarson, Mandralis, Novati, Koumoutsakos, and
  Dabiri]{gunnarson2021}
P.~Gunnarson, I.~Mandralis, G.~Novati, P.~Koumoutsakos, and J.~O. Dabiri.
\newblock Learning efficient navigation in vortical flow fields.
\newblock \emph{Nature communications}, 12\penalty0 (1):\penalty0 7143, 2021.
\newblock URL \url{https://doi.org/10.1038/s41467-021-27015-y}.

\bibitem[Fan et~al.(2020)Fan, Yang, Wang, Triantafyllou, and
  Karniadakis]{fan2020}
D.~Fan, L.~Yang, Z.~Wang, M.~S. Triantafyllou, and G.~E. Karniadakis.
\newblock Reinforcement learning for bluff body active flow control in
  experiments and simulations.
\newblock \emph{Proceedings of the National Academy of Sciences}, 117\penalty0
  (42):\penalty0 26091--26098, 2020.
\newblock URL \url{https://www.pnas.org/doi/abs/10.1073/pnas.2004939117}.

\bibitem[Nair and Goza(2023)]{nair2023}
N.~J. Nair and A.~Goza.
\newblock Bio-inspired variable-stiffness flaps for hybrid flow control, tuned
  via reinforcement learning.
\newblock \emph{Journal of Fluid Mechanics}, 956:\penalty0 R4, 2023.
\newblock URL \url{https://doi.org/10.1017/jfm.2023.28}.

\bibitem[Xia et~al.(2024)Xia, Zhang, Kerrigan, and Rigas]{xia2024}
C.~Xia, J.~Zhang, E.~C. Kerrigan, and G.~Rigas.
\newblock Active flow control for bluff body drag reduction using reinforcement
  learning with partial measurements.
\newblock \emph{Journal of Fluid Mechanics}, 981, 2024.
\newblock URL \url{https://doi.org/10.1017/jfm.2024.69}.

\bibitem[Novati et~al.(2021)Novati, de~Laroussilhe, and
  Koumoutsakos]{novati2021}
G.~Novati, H.~L. de~Laroussilhe, and P.~Koumoutsakos.
\newblock Automating turbulence modelling by multi-agent reinforcement
  learning.
\newblock \emph{Nature Machine Intelligence}, 3\penalty0 (1):\penalty0 87--96,
  2021.
\newblock URL \url{http://dx.doi.org/10.1038/s42256-020-00272-0}.

\bibitem[Bae and Koumoutsakos(2022)]{bae2022}
H.~J. Bae and P.~Koumoutsakos.
\newblock Scientific multi-agent reinforcement learning for wall-models of
  turbulent flows.
\newblock \emph{Nature Communications}, 13\penalty0 (1):\penalty0 1443, 2022.
\newblock URL \url{https://doi.org/10.1038/s41467-022-28957-7}.

\bibitem[Renn and Gharib(2022)]{renn2022}
P.~I. Renn and M.~Gharib.
\newblock Machine learning for flow-informed aerodynamic control in turbulent
  wind conditions.
\newblock \emph{Communications Engineering}, 1\penalty0 (1):\penalty0 45, 2022.
\newblock URL \url{https://doi.org/10.1038/s44172-022-00046-z}.

\bibitem[Kaelbling et~al.(1998)Kaelbling, Littman, and
  Cassandra]{kaelbling1998}
L.~P. Kaelbling, M.~L. Littman, and A.~R. Cassandra.
\newblock Planning and acting in partially observable stochastic domains.
\newblock \emph{Artificial intelligence}, 101\penalty0 (1-2):\penalty0 99--134,
  1998.
\newblock URL \url{https://doi.org/10.1016/S0004-3702(98)00023-X}.

\bibitem[Kaelbling et~al.(1996)Kaelbling, Littman, and Moore]{kaelbling1996}
L.~P. Kaelbling, M.~L. Littman, and A.~W. Moore.
\newblock Reinforcement learning: A survey.
\newblock \emph{Journal of artificial intelligence research}, 4:\penalty0
  237--285, 1996.
\newblock URL \url{https://doi.org/10.1613/jair.301}.

\bibitem[Lin and Mitchell(1992)]{lin1992}
L.~Lin and T.~M. Mitchell.
\newblock \emph{Memory Approaches to Reinforcement Learning in Non-Markovian
  Domains}.
\newblock Carnegie Mellon University, 1992.
\newblock URL \url{https://dl.acm.org/doi/10.5555/865158}.

\bibitem[Mnih et~al.(2013)Mnih, Kavukcuoglu, Silver, Graves, Antonoglou,
  Wierstra, and Riedmiller]{mnih2013}
V.~Mnih, K.~Kavukcuoglu, D.~Silver, A.~Graves, I.~Antonoglou, D.~Wierstra, and
  A.~Riedmiller.
\newblock Playing atari with deep reinforcement learning, 2013.
\newblock URL \url{https://doi.org/10.48550/arXiv.1312.5602}.

\bibitem[Hausknecht and Stone(2015)]{hausknecht2015}
M.~Hausknecht and P.~Stone.
\newblock Deep recurrent q-learning for partially observable mdps.
\newblock In \emph{2015 AAAI Fall Symposium Series}, 2015.
\newblock URL \url{https://doi.org/10.48550/arXiv.1507.06527}.

\bibitem[Meng et~al.(2021)Meng, Gorbet, and Kuli{\'c}]{meng2021}
L.~Meng, R.~Gorbet, and D.~Kuli{\'c}.
\newblock Memory-based deep reinforcement learning for pomdps.
\newblock In \emph{2021 IEEE/RSJ International Conference on Intelligent Robots
  and Systems (IROS)}, pages 5619--5626. IEEE, 2021.
\newblock URL \url{https://doi.org/10.1109/IROS51168.2021.9636140}.

\bibitem[Fujimoto et~al.(2018)Fujimoto, van Hoof, and Meger]{fujimoto2018}
S.~Fujimoto, H.~van Hoof, and D.~Meger.
\newblock Addressing function approximation error in actor-critic methods.
\newblock \emph{CoRR}, abs/1802.09477, 2018.
\newblock URL \url{http://arxiv.org/abs/1802.09477}.

\bibitem[Beckers(2023)]{phdthesis}
D.~Beckers.
\newblock \emph{Fast models and reinforcement learning control of unsteady
  aerodynamics}.
\newblock PhD thesis, University of California, Los Angeles, 2023.
\newblock URL \url{https://escholarship.org/uc/item/21r4z6xk}.

\bibitem[Haarnoja et~al.(2018)Haarnoja, Zhou, Abbeel, and Levine]{haarnoja2018}
T.~Haarnoja, A.~Zhou, P.~Abbeel, and S.~Levine.
\newblock Soft actor-critic: Off-policy maximum entropy deep reinforcement
  learning with a stochastic actor, 2018.

\bibitem[Wagner(1925)]{wagner1925}
H.~Wagner.
\newblock Über die entstehung des dynamischen auftriebes von tragflügeln.
\newblock \emph{ZAMM - Journal of Applied Mathematics and Mechanics /
  Zeitschrift für Angewandte Mathematik und Mechanik}, 5\penalty0
  (1):\penalty0 17--35, 1925.
\newblock \doi{https://doi.org/10.1002/zamm.19250050103}.

\bibitem[Theodorsen(1935)]{theodorsen1935}
T.~Theodorsen.
\newblock General theory of aerodynamic instability and the mechanism of
  flutter.
\newblock Technical Report NACA-TN-496, NACA, 1935.

\bibitem[Eldredge(2022)]{eldredge2022}
J.~D. Eldredge.
\newblock A method of immersed layers on cartesian grids, with application to
  incompressible flows.
\newblock \emph{Journal of Computational Physics}, 448:\penalty0 110716, 2022.

\bibitem[Raffin et~al.(2021)Raffin, Hill, Gleave, Kanervisto, Ernestus, and
  Dormann]{stable-baselines3}
A.~Raffin, A.~Hill, A.~Gleave, A.~Kanervisto, M.~Ernestus, and N.~Dormann.
\newblock Stable-baselines3: Reliable reinforcement learning implementations.
\newblock \emph{Journal of Machine Learning Research}, 22\penalty0
  (268):\penalty0 1--8, 2021.
\newblock URL \url{http://jmlr.org/papers/v22/20-1364.html}.

\bibitem[von K{\'{a}}rm{\'{a}}n and Sears(1938)]{vonkarman1938}
T.~von K{\'{a}}rm{\'{a}}n and W.~R. Sears.
\newblock Airfoil theory for non-uniform motion.
\newblock \emph{Journal of the Aeronautical Sciences}, 5\penalty0
  (10):\penalty0 379--390, 1938.
\newblock URL \url{https://doi.org/10.2514/8.674}.

\bibitem[Neumark(1952)]{neumark1952}
S.~Neumark.
\newblock Pressure distribution on an airfoil in nonuniform motion.
\newblock \emph{Journal of the Aeronautical Sciences}, 19\penalty0
  (3):\penalty0 214--215, 1952.
\newblock URL \url{https://doi.org/10.2514/8.2219}.

\bibitem[Jones(1938)]{jones1938}
R.~T. Jones.
\newblock Operational treatment of the non-uniform lift theory in airplane
  dynamics.
\newblock Technical Report NACA-TN-667, NACA, 1938.

\bibitem[Garrick(1938)]{garrick1938}
I.~E. Garrick.
\newblock On some reciprocal relations in the theory of nonstationary flows.
\newblock Technical Report NACA-TN-629, NACA, 1938.

\end{thebibliography}
